\documentclass[useAMS,usenatbib]{mn2e}
\usepackage{graphicx}
\usepackage{subfig}
\usepackage{amsmath}
\usepackage{amssymb}
\usepackage{mathtools}
\usepackage{placeins}
\usepackage{multicol}
\usepackage[hyperindex,breaklinks=true, colorlinks, citecolor=blue]{hyperref}  

\def\reff@jnl#1{{\rm#1\/}}
\def\aj{\reff@jnl{AJ}}                  
\def\araa{\reff@jnl{ARA\&A}}            
\def\apj{\reff@jnl{ApJ}}                
\def\apjl{\reff@jnl{ApJ}}               
\def\aplett{\reff@jnl{ApJ}}             
\def\apjs{\reff@jnl{ApJS}}        
\def\ao{\reff@jnl{Appl.Optics}}         
\def\apss{\reff@jnl{Ap\&SS}}            
\def\aap{\reff@jnl{A\&A}}               
\def\aapr{\reff@jnl{A\&A~Rev.}}  
\def\aaps{\reff@jnl{A\&AS}}    
\def\azh{\reff@jnl{AZh}}                
\def\baas{\reff@jnl{BAAS}}              
\def\jrasc{\reff@jnl{JRASC}}            
\def\memras{\reff@jnl{MmRAS}}           
\def\mnras{\reff@jnl{MNRAS}}            
\def\nar{\reff@jnl{NAR}}           	
\def\pasa{\reff@jnl{PASA}}			  
\def\pra{\reff@jnl{Phys.Rev.A}}         
\def\prb{\reff@jnl{Phys.Rev.B}}   
\def\prc{\reff@jnl{Phys.Rev.C}}         
\def\prd{\reff@jnl{Phys.Rev.D}}         
\def\prl{\reff@jnl{Phys.Rev.Lett}}      
\def\pasp{\reff@jnl{PASP}}              
\def\pasj{\reff@jnl{PASJ}}              
\def\qjras{\reff@jnl{QJRAS}}            
\def\skytel{\reff@jnl{S\&T}}            
\def\solphys{\reff@jnl{Solar~Phys.}}    
\def\sovast{\reff@jnl{Soviet~Ast.}}     
\def\ssr{\reff@jnl{Space~Sci.Rev.}}     
\def\zap{\reff@jnl{ZAp}}                
\def\nat{\reff@jnl{Nature}}             

\title[AME constraints in LDN\,1622 with the GBT]
  {Observations of Free-Free and Anomalous Microwave Emission from LDN\,1622 with the 100\,m Green Bank Telescope}
\author[S. E. Harper et al.]
  {S.E.~Harper,$^1$  C.~Dickinson,$^1$ K.~Cleary,$^2$ \\
  $^1$Jodrell Bank Centre for Astrophysics, Alan Turing Building, School of Physics and Astronomy, The University of Manchester, \\ Oxford Road, Manchester, M13 9PL, U.K.\\
  $^2$Cahill Center for Astronomy and Astrophysics, California Institute of Technology, Pasadena, CA 91125, USA}
\date{Accepted xxxx xx xx. Received xxxx xx xx}
\pagerange{\pageref{firstpage}--\pageref{lastpage}}
\pubyear{xxxx}

\begin{document}

\maketitle

\begin{abstract}
LDN\,1622 has previously been identified as a possible strong source
of dust-correlated Anomalous Microwave Emission (AME). Previous
observations were limited by resolution meaning that the radio
emission could not be compared with current generation high-resolution
infrared data from \textit{Herschel}, \textit{Spitzer} or
\textit{WISE}. This \textit{Paper} presents arcminute resolution
mapping observations of LDN\,1622 at 4.85\,GHz and 13.7\,GHz using the
100\,m Robert C. Byrd Green Bank Telescope. The 4.85\,GHz map reveals
a corona of free-free emission enclosing LDN\,1622 that traces the
photo-dissociation region of the cloud. The brightest peaks of the
4.85\,GHz map are found to be within $\approx$\,10\,\% agreement with
the expected free-free predicted by SHASSA H$\alpha$ data of
LDN\,1622. At 13.7\,GHz the AME flux density was found to be
7.0\,$\pm$\,1.4\,mJy and evidence is presented for a rising spectrum
between 13.7\,GHz and 31\,GHz. The spinning dust model of AME is found
to naturally account for the flux seen at 13.7\,GHz. Correlations
between the diffuse 13.7\,GHz emission and the diffuse mid-infrared
emission are used to further demonstrate that the emission originating
from LDN\,1622 at 13.7\,GHz is described by the spinning dust model.

\end{abstract}


\section{Introduction}

Dust-correlated anomalous microwave emission (AME) has been observed
in the frequency range 10\,$-$\,60\,GHz and has a spectrum distinct
from other sources of Galactic emission in the same range, such as
free-free, synchrotron and the cosmic microwave background (CMB)
\citep{Kogut1996, Leitch1997, DeOliveira2004,Gold2011}. Evidence of
AME has been found in molecular clouds
\citep{Watson2005,Casassus2008,PlanckEarlyXX}, HII regions
\citep{Dickinson2009, Tibbs2012}, Lynds dark clouds
\citep{Casassus2006, Scaife2009}, large scale diffuse Galactic dust
\citep{Peel2012} and in one external galaxy \citep{Murphy2010}. As AME
is present in many different environments, it may become an important
new tool for studying the interstellar medium (ISM).

There have been several proposed mechanisms for AME. The spinning dust
model is currently the favoured explanation for AME
\citep{Draine1998a, Draine1998b}. The other possible explanations for
AME include: free-free emission from shock-heated gas \citep{Leitch1997};
flat spectrum synchrotron emission \citep{Bennett2003} and magnetic
dipole emission from dust grains \citep{Draine1999}. For the purposes
of this \textit{Paper} the spinning dust model is assumed to be the
origin of AME. The spinning dust model proposes that the rapid
rotation of small dust grains with electric dipoles generates AME
\citep{Draine1998a, Draine1998b, AliHaimoud2009, Hoang2010,
  Ysard2010}. Dust grains in the interstellar medium can be broadly
separated into big grains (BGs; $> 0.1 \mu$m), very small grains
(VSGs; $< 0.1 \mu$m) and disc-like molecules known as Poly-Aromatic
Hydrocarbons (PAHs; as small as tens of atoms) \citep{Desert1990,
  Li2001}. VSGs generate mid-infrared (MIR) continuum emission,
whereas PAHs are observed as several MIR emission lines between
1\,$-$\,12$\mu$m \citep{Tielens2008}. The spinning dust model assumes
that only the VSGs and PAHs contribute to AME. Therefore strong
morphological correlations are expected between AME and mid-infrared
emission.

Recent \textit{Planck} observations have identified numerous possible
AME sources within the Galaxy that can be described by the spinning
dust model. The Perseus and $\rho$-Ophiuchus molecular clouds were
identified as the two sources that provide strong evidence in support
of the spinning dust model \citep{PlanckEarlyXX,
  PlanckIntXV}. However, it is still unknown whether PAHs or VSGs
contribute the most towards the generation of AME and whether it is
pervasive throughout the ISM. By observing AME at resolutions
comparable to current infrared space observatories, such as
\textit{WISE}, {\it Spitzer} and {\it Herschel}, it should be possible
to determine the dust grain population generating AME.

A full-width half-maximum (FWHM) resolution of 1\,arcmin or smaller
would be required to compare radio observations with infrared
(IR). These resolutions are easily obtainable using interferometric
radio observations. However, AME has a largely extended and diffuse
morphology and interferometers are only sensitive to particular
angular scales. Typically if the angular size of a source extends
beyond the synthesised beam resolution of the interferometer, the flux
from the larger angular scales is lost
\citep{ThompsonMoranSwenson2007}. The sensitivity of interferometers
to extended emission can be improved by using models of the u-v
coverage but it not possible to recover information from unobserved
scales. Single-dish radio observations are able to measure all the
flux from a source. This makes single-dish observations ideal for
measuring diffuse and extended sources. A difficulty with single-dish
observations is that they require extremely large apertures in order
to achieve resolutions comparable to the IR observatories. The 100\,m
Robert C. Byrd Green Bank Telescope (GBT) is one of the few radio
telescopes with a dish large enough to reach FWHM resolutions smaller
than $\approx$1\,arcmin.

Lynds Dark Cloud LDN\,1622 is a starless cometary cloud in the
vicinity of the Orion B molecular cloud, \textit{North-East} of
Barnard's Loop \citep{Maddalena1986}. The \textit{West} and
\textit{South} sides of LDN\,1622 are traced by a bright H$\alpha$ corona
that is clearly visible in the SHASSA images \citep{Gaustad2001} and
is illuminated by the nearby Ori OB1b stellar association
\citep{Kun2008}. The bright H$\alpha$ corona may also trace the
Photon-Dominated Region (PDR) around LDN\,1622. The Cosmic Background
Interferometer (CBI) observed LDN\,1622 at 31\,GHz with an 8\,arcmin
synthesised beam and found a strong correlation between the emissions
from the PDR seen by CBI 31\,GHz and IRAS\,12\,$\mu$m
\citep{Casassus2006}. The IR correlations seen by CBI were attributed to
heated VSGs within the PDR generating AME. Earlier observations of
LDN\,1622 over 5\,$-$\,10\,GHz with the 140\,ft Green Bank Telescope also
identified LDN\,1622 as possibly one of the first compact sources of AME
\citep{Finkbeiner2002}.  Evidence for AME was also found by the CBI at
31\,GHz in the nearby cloud LDN\,1621 (L1621), which is approximately
30\,arcmin \textit{North} of LDN\,1622 \citep{Dickinson2010}. The
proximity of LDN\,1621 to LDN\,1622 suggests they may share similar grain
populations and therefore if AME is found in one cloud it may be
present in the other.

This {\it Paper} presents new observations of LDN\,1622 at 4.85\,GHz
(C-band) and 13.7\,GHz (Ku-band) taken with the GBT. These are the
first extended mapping of a Lynds dark cloud with high resolution
radio observations from a single-dish radio telescope.  This
\textit{Paper} is structured as follows: Section \ref{sec:obs}
discusses the observations, data processing and ancillary
data. Section \ref{sec:results} describes GBT maps of LDN\,1622. Section
\ref{sec:discuss} compares the emissions seen at 4.85\,GHz and
13.7\,GHz by the GBT to archival and ancillary observations of
LDN\,1622. Finally, Section \ref{sec:conclusion} summarises the findings
in this \textit{Paper}.

\section{Observations and Data Processing} \label{sec:obs}

\subsection{GBT Observations}

The GBT is a 100\,m diameter fully-steerable single-dish radio
telescope that operates at frequencies less than 100\,GHz. The
observations were taken over several days in January 2007 and one
extra observation was made the following year in January 2008. The
observed time on source for C-band was 5\,hours and for Ku-band
7.5\,hours. 

Observations were made at two frequencies: 4.85\,GHz (C-band) and
13.7\,GHz (Ku-band). Both C-band and Ku-band observations used the
Digital Continuum Receiver (DCR) on the GBT. The observations used all
the 16 frequency channels available to the DCR for each frequency,
this provided a bandwidth of 1.28\,GHz for C-band and 3.5\,GHz for
Ku-band. The sample integration time was 0.2\,seconds. The C-band
receiver has a single beam with a Full-Width Half-Max (FWHM)
resolution of 2.6\,arcmin at 4.85\,GHz. At Ku-band the GBT has two
receivers with independent beams, both with a FWHM resolution at
13.7\,GHz of 55\,arcsec. One receiver is located in the central focal
position of the GBT and will be referred to in this \textit{Paper} as
the central beam. The other Ku-band receiver is displaced by
330\,arcsec in the cross-elevation direction and will be referred to
as the off-centre beam. Both C-band and Ku-band operated in dual
polarisation mode, measuring left (LL) and right (RR) circularly
polarised radiation simultaneously. It was assumed that the emission
from LDN\,1622 would have negligible circular polarisation, therefore
total intensity could be obtained by averaging together the LL and RR
data.

On-The-Fly (OTF) mapping was used to scan across the regions in Right
Ascension (R.A.) and Declination (Decl.), at a speed on the sky of
1\,arcmin\,s$^{-1}$. This resulted in a series of nested scans across both
regions. The C-band observations mapped a region of $35
\,\mathrm{arcmin} \times 35\,\mathrm{arcmin}$ centred upon LDN\,1622 at
R.A. $=\mathrm{5^h 54^m 29^s}$, Decl. $=\mathrm{1^\circ 45'
36''}$. The Ku-band observations mapped a $12 \,\mathrm{arcmin} \times
12\,\mathrm{arcmin}$ sub-region of LDN\,1622 centred upon
R.A. $=\mathrm{5^h 54^m 16^s}$ and Decl. $=\mathrm{1^\circ 49'
52''}$. In order to obtain the desired total integration time per beam
in the final map, multiple scans of the same region were made. By
observing in this way using OTF mapping, instead of integrating each
beam using a pointed observation, the effect of long timescale
correlated {\it 1/f} noise caused by the atmosphere or receivers could
be minimised. 

The theoretical thermal noise of the receiver was calculated using:
\begin{equation}
  \left( \frac{\sigma}{\mathrm{mK}} \right) = \frac{44.2}{\mu}\frac {\left( \frac{T_{sys}}{\mathrm{K}} \right)}{ \sqrt{ \left( \frac{\Delta \nu}{\mathrm{GHz}} \right) \left( \frac{\tau}{\mathrm{sec}} \right) } } ,
\label{eqn:thermal}
\end{equation} 
where $T_{sys}$ is the measured system temperature, $\Delta \nu$ is
the bandwidth of the receiver, $\tau$ is the sample integration time
and $\mu$ is the aperture efficiency. The aperture efficiency for
C-band was 75\,\% and for Ku-band 72\,\%. The theoretical noise levels
were 2.34\,mK and 2.1\,mK for C-band and Ku-band respectively. The
measured median noise level for all the C-band and Ku-band scans were
16.9\,mK and 13.0\,mK. Comparing the median noise level to the
theoretical noise level for C-band and Ku-band revealed that the
measured noise was 7.2 and 6.2 times higher respectively. The higher
than expected noise for both frequencies was due to \textit{1/f} noise
contamination in the data. Reducing the effect of \textit{1/f} noise
on the data is discussed in the following Section.

\subsection{Calibration and Processing}


The internal noise diode of the receiver, with a known equivalent
antenna temperature for each receiver ($T_{cal}$), was used to
calibrate the time-ordered-data (TOD) into units of brightness
temperature using the following equation:
\begin{equation}
T_{b} = \frac{T_{cal}}{2} \frac{V_{on} + V_{off}}{\langle V_{on} -
  V_{off} \rangle} - \frac{T_{cal}}{2}.
\label{eqn:caldiode}
\end{equation} 
The noise calibration diode was injected into every other sample in
the TOD at a frequency of 5\,Hz. $V_{on}$ describes the receiver
voltage when the noise diode was on and $V_{off}$ describes the
receiver voltage when the noise diode was off.

To check the diode calibration, the flux densities of 3C161, 3C48 and
3C147 were measured at the beginning and half-way through each
observing session. The measured flux densities based on the diode
calibration were then compared to the predicted flux densities from
the source models described in \citet{Ott1994}. The flux densities of
3C161, 3C48 and 3C147 measured at C-band and Ku-band after calibrating
with the diode were all found to be within 5\,\% of the predicted flux
densities.


The opacity of the atmosphere at radio wavelengths is dependent on the
water vapour content of the air and the elevation of the
observation. Over the several days of observing the relative humidity
varied considerably, between 0\,\% and 90\,\%, and the elevation of
the observations varied by 10\,degrees. Measurements of zenith opacity
from weather stations near to the GBT were used to estimate the effect
of atmospheric opacity on the data. The maximum attenuation was
estimated to be $<$0.05\,\% and $<$2.6\,\% for C-band and Ku-band. As
these corrections are less than the flux density calibration accuracy,
no atmospheric absorption corrections were applied to either C-band or
Ku-band.


The emission from ground-based radar and geostationary TV satellites
peak at around the C-band and Ku-band frequencies. These terrestrial
sources of radio emission are considered radio frequency interference
(RFI) and completely dominate astronomical sources. Several scans were
found to be RFI contaminated. These scans were identified by comparing
the peak signal in each scan to the thermal noise limit of the
receiver. Scans containing any source with a brightness greater than
20 times the receiver noise were flagged as RFI contaminated and
removed.


Time correlated noise within the TOD caused by the gain fluctuations
in the receiver amplifiers or the instability of atmospheric water
vapour can increase the effective noise of the TOD to be many times
higher than the receiver thermal noise limit. This type of noise is
known as {\it 1/f} noise due to the effect it has on the shape of the
TOD power spectrum. The spectrum of \textit{1/f} can be approximated
by a power-law \citep{Seiffert2002}:
\begin{equation}
  P_{\nu} = \sigma_{w}^{2} \left[ 1 +  \left( \frac{\nu_{knee}}{\nu} \right) ^{\alpha} \right],
  \label{eqn:fmodel}
\end{equation}
where $\sigma_{w}$ describes the thermal noise, $\nu_{knee}$ is the
knee frequency and $\alpha$ is the spectral index of the \textit{1/f}
noise. At the knee frequency the TOD thermal noise and low-frequency
\textit{1/f} noise contribute equally to the spectral density of the
TOD. The spectral index describes the power of the \textit{1/f} noise,
with typical values range between 1 and 2.

When OTF mapping continuum sources, {\it 1/f} noise can cause stripes
in the scan directions of the final map and obscure the astronomical
signal. This is a limitation of single-dish radio observations however
there are methods for mitigating {\it 1/f} noise both during the
observations and when data processing.

During the observations, the effect of \textit{1/f} noise on the
astronomical signal was reduced by slewing the telescope as fast as
possible. The data in each scan can be assumed to have \textit{1/f}
noise contributions from the receiver and the atmosphere. The noise of
each scan was assumed to have a fixed knee frequency.  By slewing the
telescope faster, the time for a scan can be made shorter than the
timescale of the \textit{1/f} noise variations. The limit on how much
\textit{1/f} noise can be removed in this way is a balance between
the knee frequency of the noise and the physical constraints of the
telescope. In the case of these observations the GBT could not slew
fast enough to remove all the \textit{1/f} noise contamination.

During the data processing stage scans containing high levels of
\textit{1/f} noise contamination were removed. The ratio between the
variance of pairs of data separated by two different lags was used as
the metric to determine the \textit{1/f} contamination of a scan. The
variance of the difference between the TOD and the TOD lagged by
$\tau$ samples ($d_{\tau + i} - d_i$) was defined as,
\begin{equation}
  \sigma_{\tau}^2 = \sigma^2(\{d_{\tau} - d_{0},d_{\tau + 1} - d_{1}, \ldots ,  d_{\tau + i} - d_{i} \} ) .
\end{equation}
Differencing the TOD separated by a lag of 1\,sample, $\tau =
0.2$\,seconds, results in a set of data with a variance
equal to the white-noise limit of the TOD but increased by a factor of
$\sqrt{2}$. The white-noise limit of the TOD can then be compared with
the variance of the TOD differenced with different lags. As the lag
time increases, more \textit{1/f} noise will contaminate the
differenced TOD. Therefore the ratio between the differenced TOD
variance and white-noise limit increases. The variance of the
differenced TOD lagged by a $\tau=1$\,second (5\,samples) was chosen
to compare with the white-noise limit of the TOD. The lag time was
chosen to be slightly shorter than the timescale of the median knee
frequency of the data, $\gtrsim$0.7\,Hz. This ensured that only scans
with a large \textit{1/f} noise contribution were filtered.

The ratio between the lagged differenced TOD and the white-noise
limit was measured for each scan as
\begin{equation}
  \label{eqn:ratio}
  R = \frac{\sigma_{\tau}}{\sigma_{W}} .
\end{equation}
The metric $R$ defines the ratio of the lagged differenced TOD
variance ($\sigma_{\tau}$) and the white noise limit ($\sigma_W$). If
a scan exceeded a cut-off value for $R$ it was discarded. Many
different cut-off values for $R$ were used to generate difference maps
using jack-knifes of the data. The optimal cut-off for $R$ was
determined when the noise in the difference maps was at a minimum.  A
value of 2.3 for $R$ was found to minimise the noise in the difference
maps at both frequencies. For C-band and Ku-band, 50\,\% and 14\,\% of
the scans exceeded the $R=$2.3 maximum cut-off noise variance ratio and
were removed.

A summary of the observations and map noise limits can be found in
Table \ref{tab:receiver}.

\begin{figure*}
  \centering
  \subfloat{\includegraphics[width=0.3\textwidth]{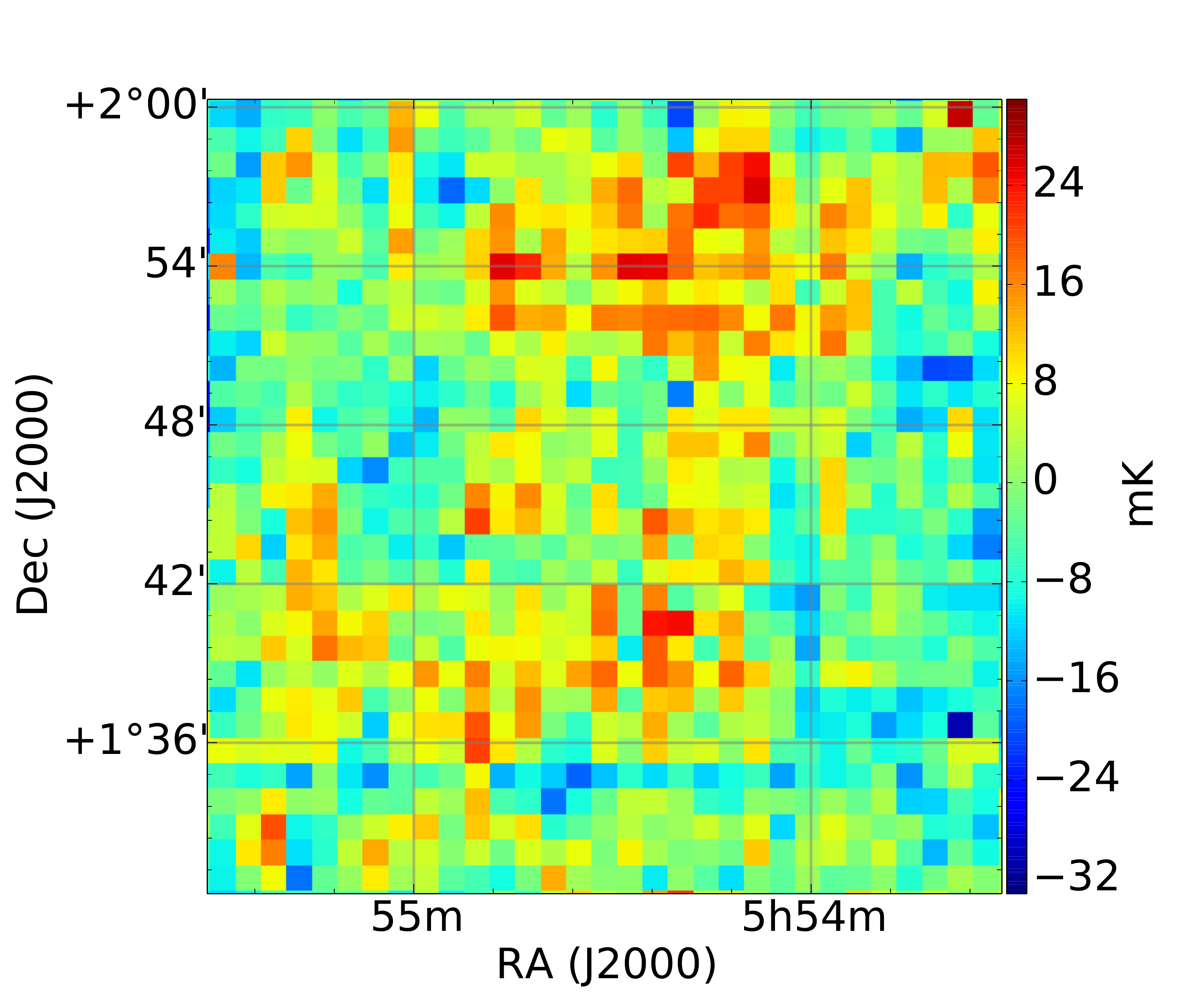}}
  \qquad
  \subfloat{\includegraphics[width=0.3\textwidth]{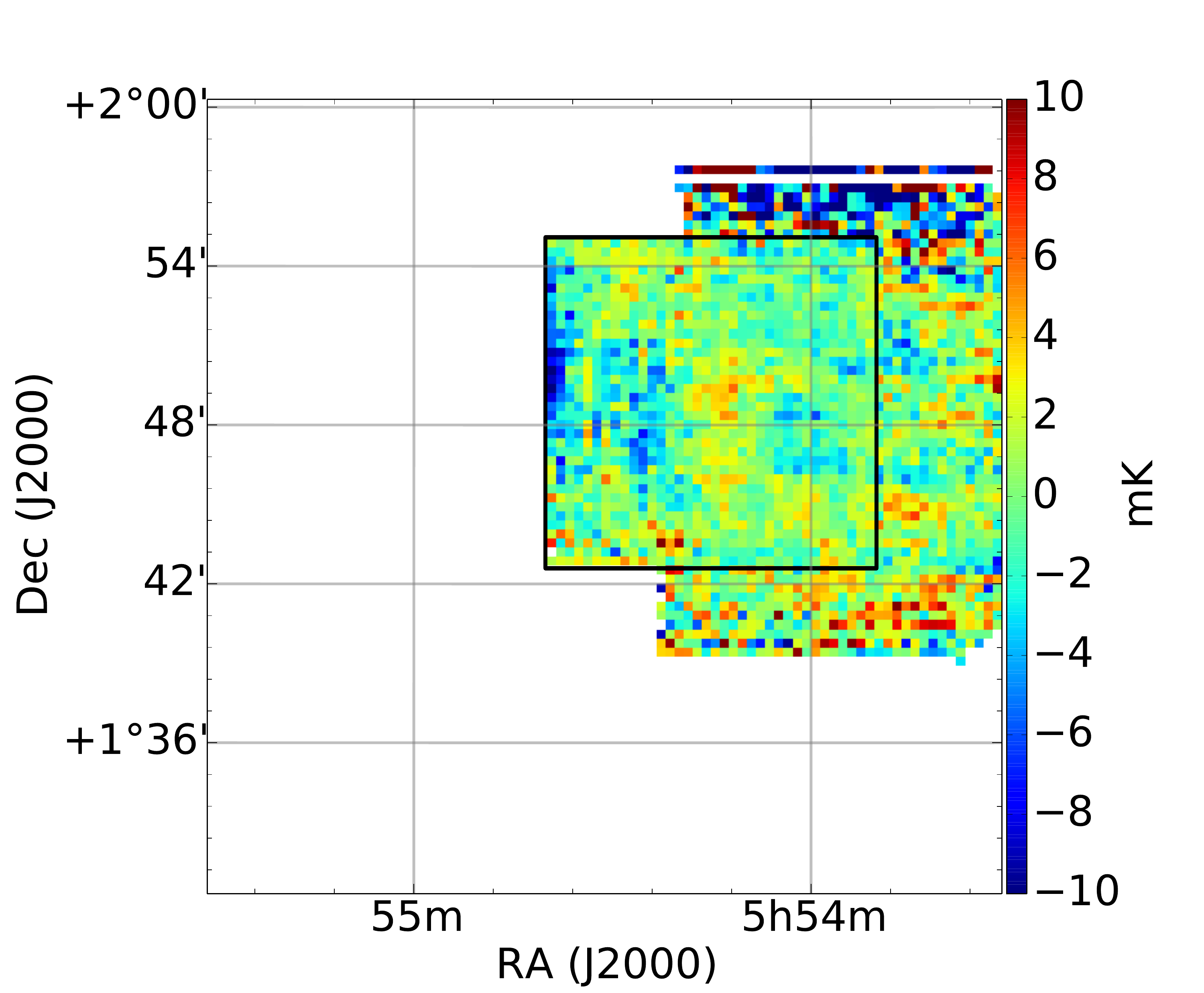}}
  \qquad
  \subfloat{\includegraphics[width=0.3\textwidth]{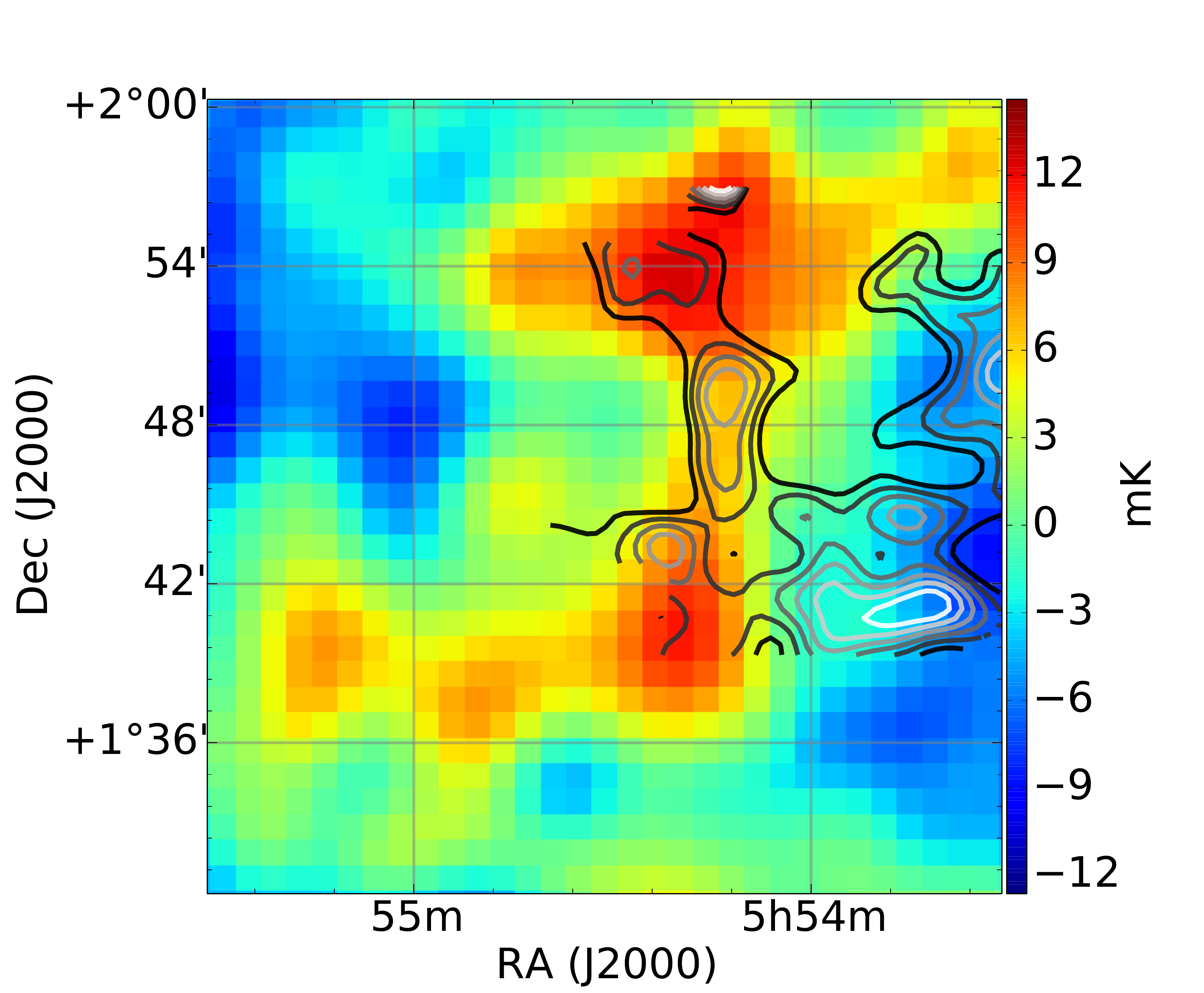}}

  \caption{Final C-band and Ku-band maps as generated by the ML
  map-making method. The images are centred at R.A. $=\mathrm{5^h54^m
  29^s}$, Decl. $=\mathrm{1^\circ 45' 36''}$ with dimensions
  30\,arcmin$\times$30\,arcmin.  \textit{Left}: C-band map at the
  original 2.6\,arcmin FWHM resolution. \textit{Centre}: Ku-band map
  at the original 55\,arcsec FWHM resolution. The black box marks the
  extent of the area observed by the central beam. \textit{Right}:
  Overlay of smoothed Ku-band map contours on the smoothed C-band
  map. The contour levels are 0.1, 0.45, 0.8, 1.15, 1.5 and
  1.85\,mJy\,beam$^{-1}$ for a 1.2\,arcmin beam.}
  \label{fig:GBTMaps}
\end{figure*}

\subsection{Map-Making}

The maximum-likelihood (ML) map-making technique was used as the final
step in reducing \textit{1/f} in the final maps. ML map-making was
originally developed for CMB experiments as a way of recovering the
weak CMB signal from {\it 1/f} noise dominated observations
\citep{Borrill1999,Natoli2001,Dore2001}. ML map-making is the ideal
method to use for these observations as for many scans the emission
from LDN\,1622 was dominated by \textit{1/f} noise.

The advantage of ML map-making is that it uses an estimated full
covariance matrix of the noise to find an optimal solution for the
signal in the TOD. The estimated covariance matrix accounts for both
the instrumental and atmospheric contributions to the \textit{1/f}
noise. The work in this \textit{Paper} used an independently
developed ML map-maker. The ML map-makers ability to recover signal
from a \textit{1/f} contaminated set of TOD was tested using
simulations. A summary of the ML map-making technique, this
implementation and simulations are described in Appendix
\ref{sec:AppA}.

\subsection{Description of Ancillary Data}

In this \textit{Paper} the GBT data at C-band and Ku-band were
compared with existing data at radio, IR and optical
frequencies. Section \ref{sec:compmfd} discusses the ancillary data
and its relationship to the emissions at C-band and Ku-band in more detail.

Ancillary interferometric radio data at 15.7\,GHz was provided by the
Arcminute Microkelvin Imager (Yvette Perrot; private communication)
\citep[AMI]{Zwart2008}. The AMI Small Array (SA) consists of 10 3.7\,m
diameter dishes. The AMI SA has baseline spacings between 5\,$-$\,20\,m and a
synthesised beam of 3\,arcmin.

FIR observations were obtained from publicly available archival ESA
\textit{Herschel} Space Observatory \citep{Pilbratt2010} data. The
archival data were SPIRE \citep{Griffin2010} and PACS
\citep{Poglitsch2010} photometry of Orion B at 500\,$\mathrm{\mu}$m,
350\,$\mathrm{\mu}$m, 250\,$\mathrm{\mu}$m and
160\,$\mathrm{\mu}$m. The FWHM resolutions of each band were
approximately: 35.2\,arcsec, 23.9\,arcsec, 17.6\,arcsec and
11.7\,arcsec. The SPIRE and PACS observations were made in parallel
mode with observation identification numbers OBSID:\,1342205074 and
OBSID:\,1342205075. The \textit{Herschel} maps were reduced using the
SPG v11.1.0 pipeline. For the PACS data, the level 2.5 \textsc{MadMap}
products were used. The SPIRE level 2 products with an absolute
background correction applied from \textit{Planck} data were
used. PACS 70\,$\mu$m data were also available but were not used because
of poor signal-to-noise.

MIR data were provided by the Wide-field Infrared Sky Explorer
\citep[\textit{WISE}]{Wright2010}. In this \textit{Paper} only the
22\,$\mathrm{\mu}$m and 12\,$\mathrm{\mu}$m observations were used in
order to reduce the number of point sources contaminating the field.
The FWHM resolution of the \textit{WISE} 22\,$\mathrm{\mu}$m map is
12\,arcsec and the \textit{WISE} 12\,$\mathrm{\mu}$m map is
\,7.4\,arcsec.

An archival Spitzer IRAC \citep{Fazio2004} map of LDN\,1622 at 8\,$\mu$m
was used to supplement the WISE MIR data. Point sources from the
Spitzer map were subtracted. The IRAC 8\,$\mu$m map has a FWHM
resolution of $\approx$1.98\,arcsec.

Optical H$\mathrm{\alpha}$ data of LDN\,1622 were taken from the Southern
H-Alpha Sky Survey Atlas \citep[SHASSA]{Gaustad2001}. The SHASSA
H$\alpha$ data serves as a tracer for the free-free emission seen at
C-band \citep{Dickinson2003}. The SHASSA map has a FWHM resolution of
0.8\,arcmin and can detect sources down to the level of
$\approx$2\,Rayleigh. The SHASSA continuum-subtracted maps contain
numerous stellar artefacts around poorly subtracted bright
sources. However, no significant stellar artefact contamination of the
H$\alpha$ emission observed around LDN\,1622 was visible.


 \begin{table} 
   \caption{Receiver and observational properties of the 4.85\,GHz (C-band) and 13.7\,GHz  (Ku-band) datasets. \label{tab:receiver}}
   \begin{center}
   \begin{tabular}{lcc}
   \hline
     Central Frequency (GHz)                                    & 4.85                & 13.7 	              \\ \hline
     Bandwidth (GHz)                                            & 1.28                & 3.5                   \\
     No. Beams                                                  & 1                   & 2                     \\
     Back end                                                   & DCR                 & DCR                   \\
     Observing mode                                             & On-The-Fly          & On-The-Fly            \\
     Hours Observed (Hours)                                     & 5                   & 7.5                   \\
     Beam (FWHM)                                                & 2.6\,arcmin         & 55\,arcsec            \\
     Smoothed Beam (FWHM)                                       & 3\,arcmin           & 1.2\,arcmin           \\
     Sky scan speed (arcmin\,s$^{-1}$)                           & 1                   & 0.5                   \\
     Data Flagged (percent)                                     & 50                  & 14                    \\     
     Map Noise (mJy\,beam$\mathrm{^{-1}}$)                        & 4.3                 & 5.5                   \\ 
     Confusion limit\footnotemark[1] (mJy\,beam$\mathrm{^{-1}}$)  & $\approx$\,0.45\,   & $\approx$\,0.03       \\
     Map Noise (mK)                                             & 2.5                 & 2.5                   \\ 
     Confusion limit\footnotemark[1] (mK)                                       & $\approx$\,0.9\,    & $\approx$\,0.05       \\
     Field Width (arcmin)                                       & 35                  & 12                    \\
  \hline
   \end{tabular}
   \end{center}
 \end{table}

 \footnotetext[1]{Confusion limit calculated using
 \begin{equation}
   \mathrm{\frac{\sigma_c}{mJy \, beam^{-1}}} = 0.2 \left( \mathrm{
     \frac{\nu}{GHz} } \right)^{-0.7} \left( \mathrm{ \frac{\theta}{arcmin} } \right)^2,
 \end{equation}
 which is a parametrised form of the calculation found in
 \citet{Condon1974}.}


\section{Results} \label{sec:results}

\subsection{GBT Maps}

Fig. \ref{fig:GBTMaps} presents the final C-band and Ku-band ML maps,
both unsmoothed and smoothed. The unsmoothed C-band map has a FWHM
resolution of 2.6\,arcmin. The map pixel size for the C-band map is
57.6\,arcsec ensuring 2.7\,pixels across a beam. The pixel size was
chosen to maintain Nyquist sampling across the field.  The noise in
the C-band map was estimated from jack-knifing RR and LL polarisation
maps. The jack-knife map should contain only the white-noise and
residual receiver {\it 1/f} noise. In order to avoid the higher noise
towards the edge of the C-band map, the noise was estimated from a
20\,arcmin diameter aperture in the centre of a difference map
generated from jack-knife maps of the data. The pixel noise within the
aperture was measured to be 2.5\,mK or 4.3\,mJy\,beam$^{-1}$.

\begin{figure*}
  \centering  
  \includegraphics[width=1.1\textwidth,trim=50 50 40 50]{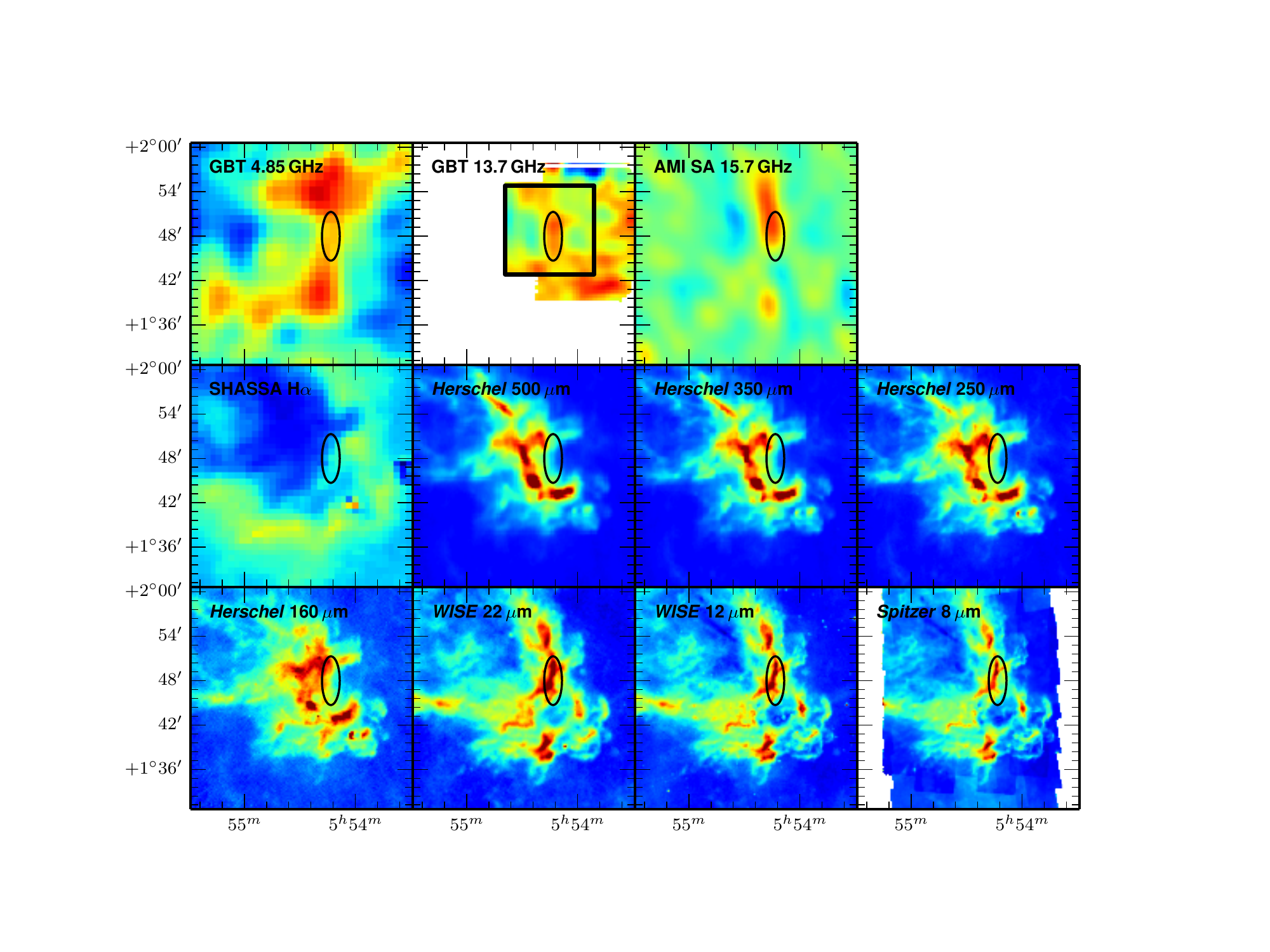}
  \caption{Maps of LDN\,1622 from GBT C-band and Ku-band observations and
  ancillary data. Each image is centred at R.A. $=\mathrm{5^h54^m
  29^s}$, Decl. $=\mathrm{1^\circ 45' 36''}$ with dimensions of
  30\,arcmin$\times$30\,arcmin. The elliptical aperture in each map
  marks the peak in Ku-band emission. A box surrounds the region
  observed by the central Ku-band beam in the Ku-band map. From the
  \textit{top left} to the \textit{bottom right} the images are: GBT
  C-band 4.85\,GHz smoothed to a 3\,arcmin beam and Ku-band 13.7\,GHz
  smoothed to a 1.2\,arcmin beam; AMI 15.7\,GHz Small Array; SHASSA
  H$\alpha$; \textit{Herschel} SPIRE 500\,$\mathrm{\mu}$m,
  350\,$\mathrm{\mu}$m and 250\,$\mathrm{\mu}$m; \textit{Herschel}
  PACS 160\,$\mathrm{\mu}$m; \textit{WISE} 22\,$\mathrm{\mu}$m and
  12\,$\mathrm{\mu}$m; \textit{Spitzer} IRAC 8\,$\mu$m. }

  \label{fig:ObsGrid}
\end{figure*}

The Ku-band map is shown alongside the C-band map in
Fig. \ref{fig:GBTMaps}. The original resolution of the Ku-band map was
55\,arcsec FWHM and the map has 2.7\,pixels per beam FWHM with a pixel
size of 20.5\,arcsec. As with the C-band map, the pixel size was
chosen to ensure Nyquist sampling across the map. The Ku-band map
combines data from a central beam and an off-centre beam. The
off-centre beam, which is displaced by 330\,arcsec in azimuth from
focal centre, observed a larger region than the central beam as the
observations were made at several different Hour Angles.  For this
reason the Western edge of the map has lower sampling and far higher
noise than the Eastern side of the map, which was observed by both
beams. The region observed by both beams is marked in
Fig. \ref{fig:GBTMaps} and \ref{fig:ObsGrid} with a box. In order to
avoid the regions of high noise within the Ku-band map the noise in
the map was measured from a difference map, generated from jack-knifes
of the data, inside an aperture of 6\,arcmin diameter centred on the
Ku-band source marked in Fig. \ref{fig:ObsGrid}. The noise was found
to be 2.5\,mK or 5.5\,mJy\,beam$^{-1}$.

Fig. \ref{fig:GBTMaps} also shows smoothed contours of the Ku-band map
overlaid onto a smoothed C-band map. The Ku-band map was smoothed to
1.2\,arcmin FWHM resolution and the C-band map was smoothed to 3\,arcmin
FWHM resolution. The smoothing kernel for both GBT maps was chosen to
increase signal-to-noise while retaining the structure of the emission
within the field. The noise level of the smoothed C-band map is
1.7\,mK or 2.9\,mJy\,beam$^{-1}$ and the noise level of the smoothed Ku-band
map is 0.8\,mK and 1.8\,mJy\,beam$^{-1}$.

\subsection{Comparison with Multi-Frequency Data}\label{sec:compmfd}

In Fig. \ref{fig:ObsGrid} the smooth C-band and Ku-band maps are shown
alongside ancillary maps of LDN\,1622. 

The C-band free-free emission in Fig. \ref{fig:ObsGrid} can be seen to
enclose LDN\,1622 in a corona which spans from the \textit{South-East} to
the \textit{North-West} and arches towards the \textit{South-West}
corner of the map. The H$\alpha$ emission, a known tracer of free-free
emission \citep{Dickinson2003}, can also be seen to have formed a
corona of emission with a similar morphology to the corona of C-band
emission. A key feature in the C-band and H$\alpha$ maps is the
Southern bar that spans the cloud \textit{East} to \textit{West}. This
feature seems to correlate quite well between the C-band and H$\alpha$
maps. The rest of the corona shows significant differences in
morphology between the H$\alpha$ and C-band emission, which can be
mostly attributed to dust absorption of the H$\alpha$
emission. Corrections for dust absorption are discussed in more detail
in Section \ref{sec:FreeFree}.

The corona, seen as both free-free and H$\alpha$ emission, is likely
associated with warm HII gas within the photo-dissociation region
(PDR) around LDN\,1622. The PDR around LDN\,1622 is a transitional
region, where on the Western side gas and dust are heated and ionised
by far ultra-violet (FUV) radiation. Progressing Eastward the FUV flux
is absorbed and the gas and dust eventually cool into the atomic and
molecular phases \citep{Hollenbach1999}. The FIR and MIR structures
shown in the \textit{Herschel}, \textit{WISE} and \textit{Spitzer}
maps around the aperture marked in Fig. \ref{fig:ObsGrid} clearly show
the separate stages occurring within the PDR. The MIR maps are tracing
the emission from the warm VSGs, which are mixed with the ionised HII
gas, as shown by the C-band data, and are exposed to the ionising
interstellar radiation field. PDRs are environments rich in PAH
molecules that have emission lines which lie within the passbands of
the \textit{WISE} 12\,$\mu$m and \textit{Spitzer} 8\,$\mu$m maps
\citep{Tielens2008}. At longer wavelengths, such as in the FIR
\textit{Herschel} maps, the maps trace the location of cooler
dust. The cooler dust is located in clumps near the core of LDN\,1622
where the dust is shielded from the interstellar radiation field and
is in thermal equilibrium with the environment.


In the smoothed Ku-band map, an elongated structure can be seen running
\textit{North-South} through the marked aperture in
Fig. \ref{fig:ObsGrid}. The location of this aperture was also the
location of the peak in emission seen at 31\,GHz by CBI
\citep{Casassus2006}. The weak emission from the elongated structure
at Ku-band is also clearly visible in the Fig. \ref{fig:ObsGrid} AMI
15.7\,GHz SA map. Comparing the Ku-band map with the MIR maps reveals
a shared morphology in the region of the aperture. The correlations
between the MIR and Ku-band maps implies a possible common source. The
spinning dust model provides a possible explanation for the shared MIR
and Ku-band morphology as it provides a mechanism for PAH molecules to
emit at the Ku-band frequency \citep{AliHaimoud2009, Ysard2010}. A
more detailed discussion of the Ku-band emission and its correlations
with IR and radio emission can be found in Section \ref{sec:AME}.

Note that a bright young stellar object (YSO) at R.A. $=\mathrm{5^h
  54^m 24.3^s}$ and Decl. $=\mathrm{1^\circ 44' 19''}$ was removed
from the FIR and MIR maps. The YSO has been briefly discussed in
\citet{Bally2009} and is associated with the proto-stellar outflow
HH\,962.

\section{Analysis and Discussion} \label{sec:discuss}

\subsection{Free-Free Emission from LDN\,1622}\label{sec:FreeFree}

Ionised regions of HII gas generate both H$\alpha$ and free-free
continuum emission. The brightness of H$\alpha$ emission is directly
proportional to the density of the ionised hydrogen ($n_H$) and the
rate at which recombination occurs within the cloud, resulting in the
H$\alpha$ transition. The brightness of free-free emission depends on
the density of free electrons ($n_e$) and ions within a region. In HII
regions the free electrons are generated via ionisation therefore it is
expected that $n_e \approx n_H$ . Both emissions also depend on the
electron temperature of the region, which is typically $T_{e} \approx
10^4$\,K. As both free-free and H$\alpha$ emissions share a common
source, it is possible to directly relate the brightness of one
emission to the brightness of the other \citep{Dickinson2003}.

\begin{figure*}
  \centering
  \subfloat{\includegraphics[width=0.30\textwidth]{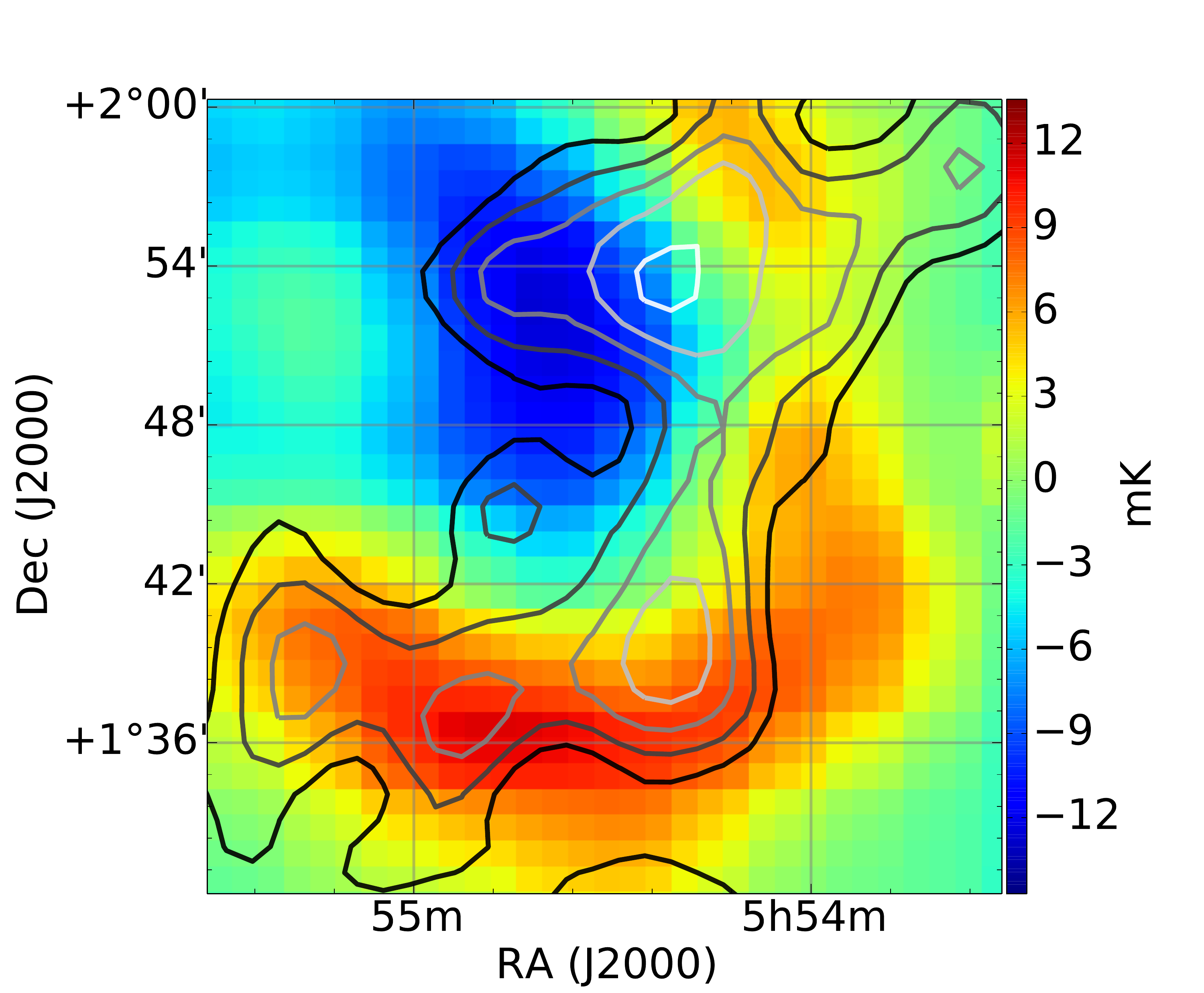}}
  \qquad
  \subfloat{\includegraphics[width=0.30\textwidth]{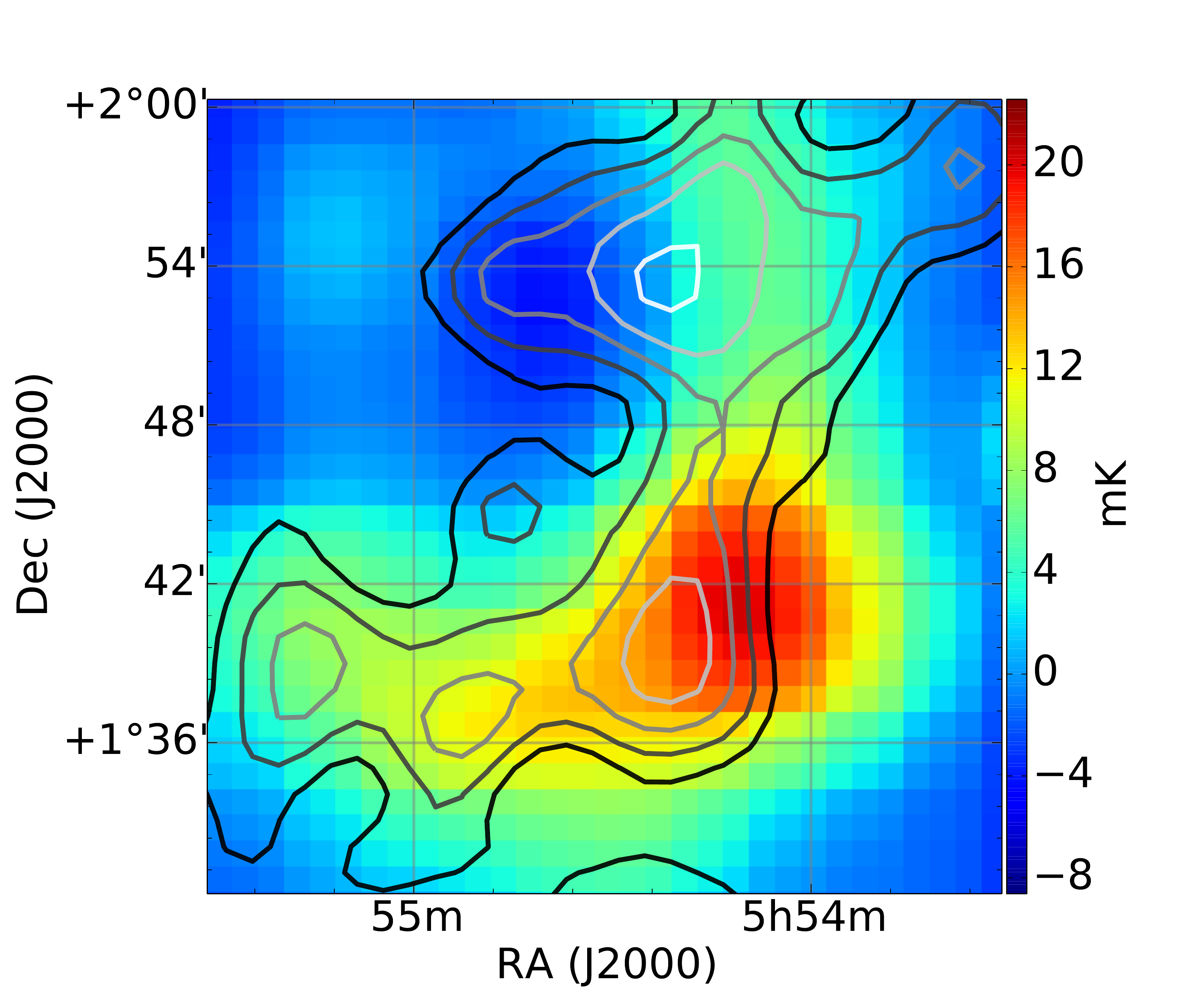}}
  \qquad
  \subfloat{\includegraphics[width=0.30\textwidth]{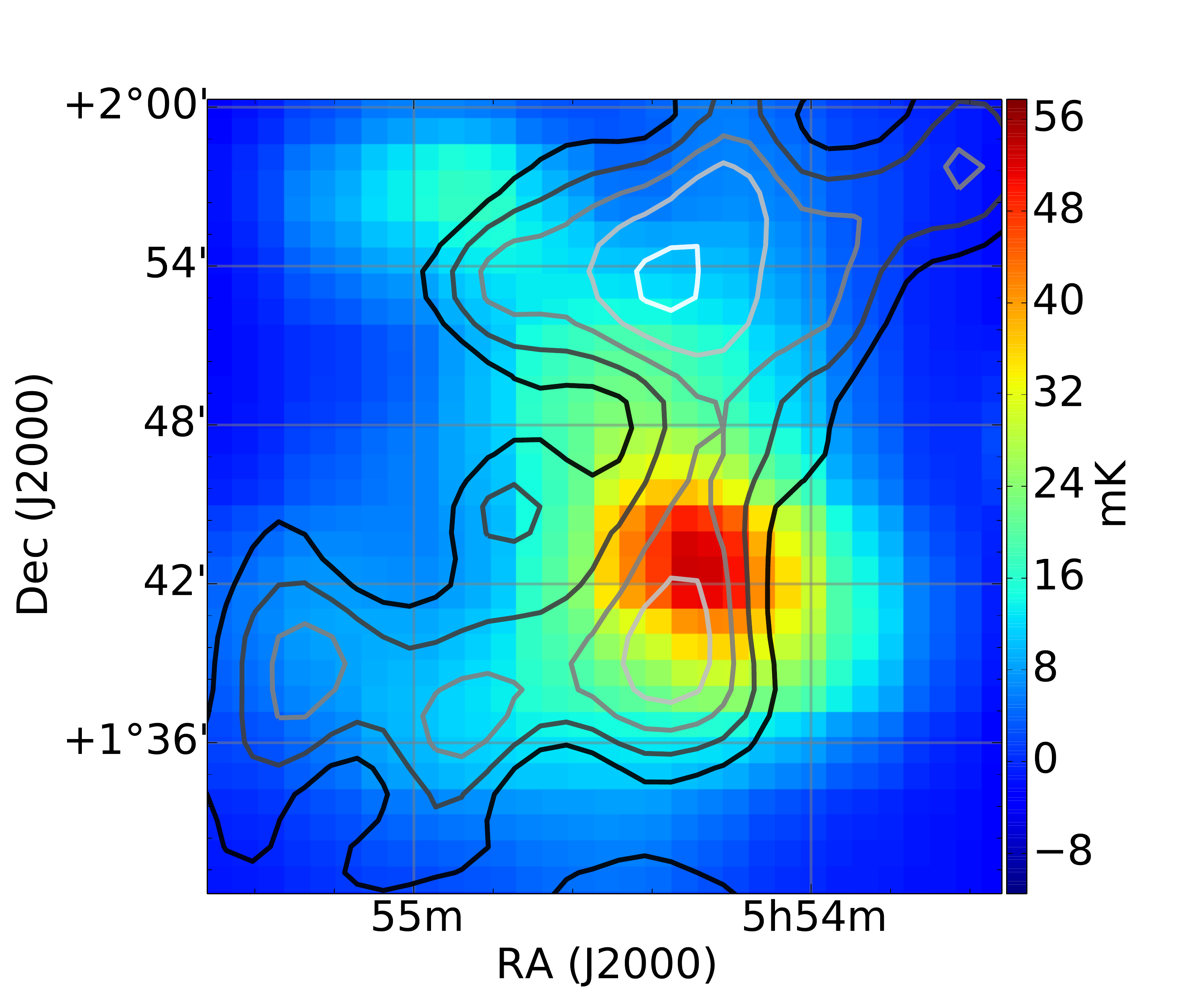}}

  \caption{Overlay of smoothed C-band contours on H$\mathrm{\alpha}$
  derived free-free maps assuming an electron temperature of
  7000\,K. The contour levels are 1, 3.75, 6.5, 9.25 and 12 in units
  of mK. The dust mixing fraction for each map from \textit{left} to
  \textit{right} is: $f = $0.1, 0.5 and 1.0.}
  \label{fig:halcorr}
\end{figure*}

In Fig. \ref{fig:ObsGrid} both the C-band and SHASSA maps show the
same corona of HII tracing the \textit{South} and \textit{West} edges
of LDN\,1622. The differences in the morphology of the radio free-free and
SHASSA H$\alpha$ emissions around LDN\,1622 are primarily due to optical
dust absorption of the H$\alpha$ emission. Correcting for H$\alpha$
dust absorption is complicated as the distribution of dust and HII gas
is unknown. In this Section the radio free-free emission is estimated
from the SHASSA map using the method outlined in \citet{Dickinson2003}
and compared with the observed C-band free-free emission. The
extinction of the H$\alpha$ by dust is estimated using a simple
radiative transfer model and optical depths measured from the
\textit{Herschel} FIR maps.

FIR \textit{Herschel} data of LDN\,1622 were used to estimate
H$\mathrm{\alpha}$ absorption, A(H$\mathrm{\alpha}$). The first step
in deriving A(H$\mathrm{\alpha}$) was to smooth all the
\textit{Herschel} maps to a common resolution. The 353\,GHz optical
depth ($\tau_{353}$) was fitted for, pixel-by-pixel, over four
\textit{Herschel} bands. Each pixel was fitted with the modified
black-body curve:
\begin{equation}
  S_{\nu} = B_{\nu}(T_D) \tau_{353} \left( \frac{\nu}{353\,\mathrm{GHz}} \right)^\beta ,
  \label{eqn:ModBB}
\end{equation}
where $\beta$, the emissivity index, the dust temperature, $T_D$, and
$\tau_{353}$ were all free parameters. $B_{\nu}$ is a black-body curve
dependent on the dust temperature.

IR optical depth was converted to hydrogen column density ($N_H$)
using:
\begin{equation}
  \tau_{353} = \sigma_{e\,353} N_H , 
\end{equation}
where $\sigma_{e\,353}$ is the mean whole sky dust opacity derived by
\textit{Planck} \citep{PlanckIntXI}. The hydrogen column density was
then related to extinction ($E(B-V)$) by assuming the relationship
$N_H/E(B-V) \approx 6.94 \times 10^{21}$\,cm$^{-2}$ \citep{PlanckIntXI}. The
H$\mathrm{\alpha}$ absorption factor was related to the measured
extinction with,
\begin{equation}
  A(\mathrm{H\alpha}) = 0.81 R(V) E(B-V)
\end{equation}
The reddening value ($R(V)$) for the ISM is typically $\approx 3.1$
\citep{Sneden1978,Schultz1975}. However, $R(V)$ is observed to
increase in denser regions of the ISM and even vary significantly
between the edge of a dark cloud and its interior with a range of
reddening values between $3.5 < R(V) < 5.5$
\citep{Vrba1985,Vrba1984}. Considering the typical reddening values of
other dark clouds and that LDN\,1622 is quite a translucent cloud, a
nominal reddening value of $R(V) = 4$ was assumed.

The expected H$\mathrm{\alpha}$ emission ($I_{em}$) was estimated from
the observed H$\mathrm{\alpha}$ emission ($I_{obs}$) by applying the
following H$\mathrm{\alpha}$ absorption model pixel-by-pixel to the SHASSA map:
\begin{equation}
  I_{obs} = \frac{I_{em}}{\tau} \int_0^{f\tau} e^{\tau'-\tau} d\tau' +  (1 - f)I_{em}
  \label{eqn:dustmdl}
\end{equation}
The H$\mathrm{\alpha}$ optical depth ($\tau$) was related to
H$\mathrm{\alpha}$ absorption by $\tau = A(\mathrm{H\alpha}) / (2.5
\log_{10}(e))$. The H$\mathrm{\alpha}$ absorption model assumed that a
certain fraction ($f$) of the HII gas and dust were mixed and in
thermal equilibrium, while the remaining fraction of HII gas was in
front of the dust and its H$\mathrm{\alpha}$ emission unabsorbed. The
model does not consider HII gas completely behind the dust within
LDN\,1622 because the H$\alpha$ emission in that case would be mostly
absorbed resulting in large uncertainties in the estimated radio
free-free. However, it is possible that some fraction of HII gas is
completely obscured by dust behind LDN\,1622.

After correcting for H$\mathrm{\alpha}$ absorption, the
H$\mathrm{\alpha}$ brightness was converted to the equivalent radio
free-free brightness using the method described in
\citet{Dickinson2003}. The derived H$\mathrm{\alpha}$ to radio
free-free conversion factor, assuming an electron temperature of $T_e
= 7000$\,K, was 0.28\,mK\,R$^{-1}$ at 4.85\,GHz. The final simulated free-free
map was smoothed to a FWHM of 3\,arcmin to match the smoothed C-band
resolution. The median background of the simulated free-free map was
subtracted to represent the effect of the map-making process removing
the zero-level of the C-band data.

Three examples of simulated free-free maps with dust-gas mixing
fractions of $f$ = 0.1, 0.5 and 1.0 are shown in
Fig. \ref{fig:halcorr} with C-band contours overlaid. At low mixing
fractions the morphology and brightness of the Southern bar of the
corona is reasonably well reconstructed by the simulated free-free
map. At higher mixing fractions a bright core of emission forms in the
\textit{West} of the simulated LDN\,1622 corona that is completely
unrepresentative of the observed C-band emission. This implies that
there is possibly very little dust-gas mixing occurring in the
\textit{South} of LDN\,1622. The morphology and brightness of the
\textit{North-West} region of the corona appears significantly
different in the predicted free-free maps to the observed C-band
emission. The brightest region of the C-band emission in the
\textit{North-West} region of the corona is offset towards the
\textit{East} of the predicted free-free emission. An explanation for
this is that much of the HII gas generating the free-free emission is
behind the dust in LDN\,1622, therefore the H$\alpha$ emission has been
completely absorbed along that line-of-sight.

The discrepancies between the observed C-band emission and simulated
free-free emission could be explained by the dust-gas mixing fraction
($f$) varying across the cloud. It is possible that changes in the
dust-gas mixing fraction is not the only explanation. The model
assumed that both the 353\,GHz dust opacity ($\sigma_{e\,353}$) and
optical reddening value ($R(V)$) were both constant across
LDN\,1622. However, IR dust opacity and reddening are both linked to
grain size distributions, which could be changing from the edge of
LDN\,1622 towards the core \citep{Weingartner2001}.

The comparison between the SHASSA H$\alpha$ and CBI 31\,GHz
observations of LDN\,1622 \citep[Fig. 9]{Casassus2006} did not show any
emission originating from the \textit{North-Western} part of the
corona that can be seen as a bright feature in the C-band map. Therefore
an upper limit of 16.7\,mK on the free-free emission was estimated
from the contours of the CBI map. This upper limit is in agreement
with the measured brightness temperatures of the C-band emission that
range between 6\,$-$\,11\,mK.

To summarise, the emission detected at C-band is seen to trace the HII
corona of LDN\,1622 and broadly agrees with the morphology of the SHASSA
H$\alpha$ map. The peak free-free brightness predicted by the
H$\alpha$ emission agrees within $\approx$10\,\% of the peak observed
C-band free-free emission for the bright H$\alpha$ ridge in the
\textit{South} of LDN\,1622 for dust mixing fractions of $f \lesssim
0.1$. Many of the differences in brightness and morphology between the
predicted and observed free-free emission can be attributed to local
variations in the dust-gas mixing fraction, reddening value or IR dust
opacity over LDN\,1622.


\subsection{AME at Ku-band}\label{sec:AME}

In the smoothed 13.7\,GHz Ku-band GBT map shown in Fig.
\ref{fig:ObsGrid}, the aperture shown is centred on the peak in Ku-band
emission at R.A. $=\mathrm{5^h54^m 13^s}$, Decl. $=\mathrm{1^\circ 47'
49''}$. The aperture has a semi-major axis of 3\,arcmin and a
semi-minor axis of 1.25\,arcmin.

Visually, Fig. \ref{fig:ObsGrid} shows that there is some
contribution of free-free emission from HII gas as well as MIR emission
from dust grains within the aperture. This Section will determine
whether the emission at Ku-band is AME or just free-free
emission. This required calculating the expected free-free emission
flux at Ku-band and assessing whether there is any significant
morphological similarities between the emissions at Ku-band and the
MIR.

Aperture photometry was used to measure the flux within the Ku-band
aperture shown in Fig. \ref{fig:ObsGrid}. The aperture size of
3\,arcmin by 1.25\,arcmin was chosen by expanding the aperture axes
from zero until the signal-to-noise ratio within the aperture was
maximised. The uncertainty and background within the Ku-band aperture
was estimated from an elliptical annulus around the aperture. The
annulus had inner and outer semi-minor axes of 2.17\,$-$\,4.68\,arcmin
and inner and outer semi-major axes of 5.2\,$-$\,11.2\,arcmin. The
annulus inner and outer axes were chosen to enclose a sufficiently
large sample of the C-band and Ku-band maps. Due to the Ku-band source
being in close proximity to a number of diffuse sources, the estimated
uncertainty on the flux density was measured from a difference map
generated from jack-knife maps of each observing day. By differencing
different days, the difference map contained a statistically similar
\textit{1/f} noise contribution from the atmosphere and receivers as
the Ku-band map. A five percent systematic uncertainty from the
calibration of the data were added in quadrature to the measured
uncertainties of the C-band and Ku-band flux densities.

The flux density within the Ku-band aperture shown in
Fig. \ref{fig:ObsGrid} was measured as 7.0\,$\pm$\,1.4\,mJy. A
free-free flux density of 0.1$\pm$2.4\,mJy was measured from the
C-band map within the same aperture. Inspection of the C-band map in
Fig. \ref{fig:ObsGrid} reveals that there must be free-free emission
originating from within the aperture. Measuring the same aperture but
using a background derived from the outside of the HII corona gives a
C-band flux density of 5.8\,$\pm$\,2.4\,mJy within the Ku-band
aperture. However, over the region covered by the Ku-band map the
free-free emission appears to be largely uniform and therefore zero
with respect to the local background.

\begin{figure}
  \centering
  \subfloat{\includegraphics[width=0.45\textwidth]{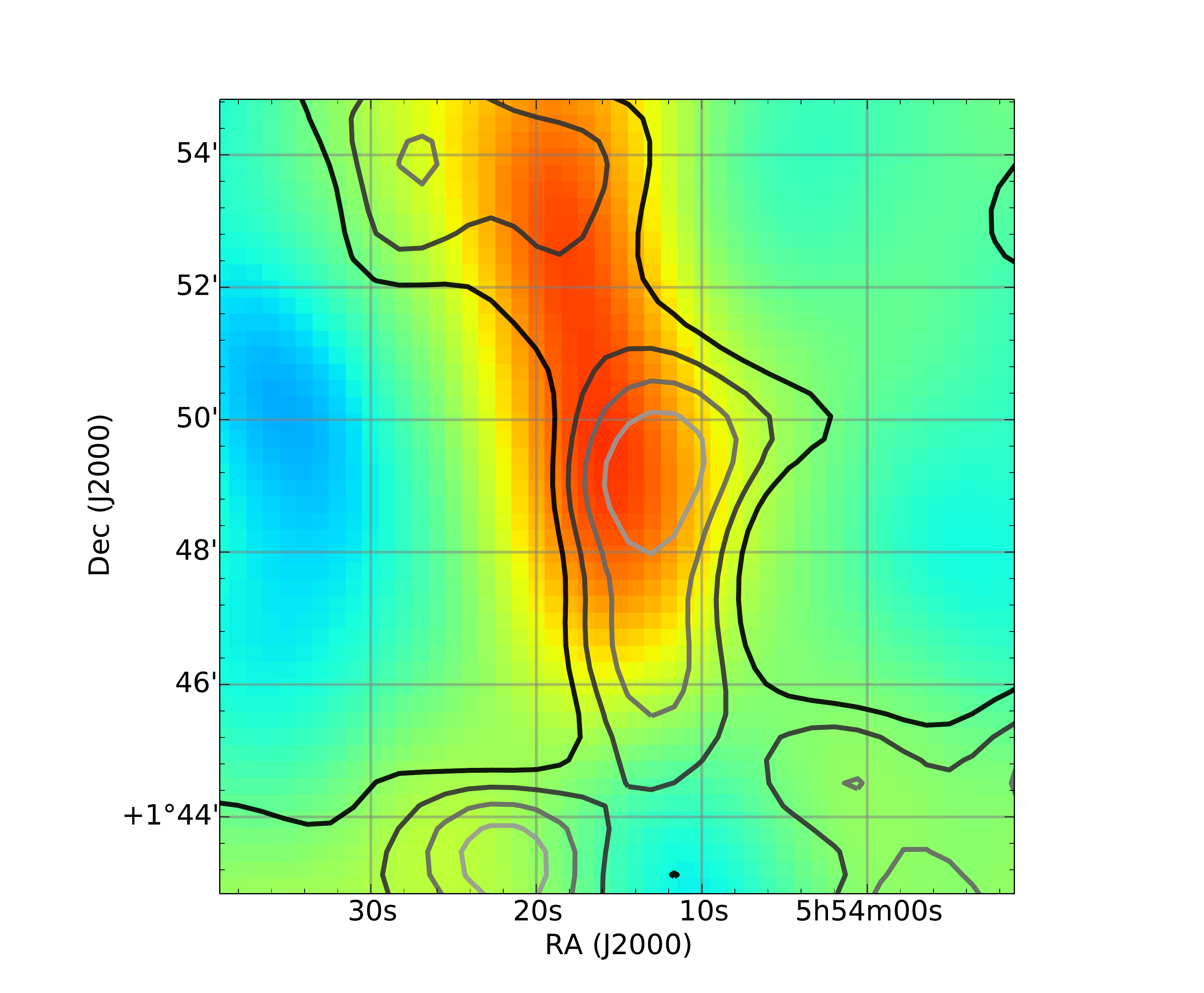}}
  \caption{Overlay of smoothed GBT Ku-band contours on AMI SA
    colourscale. The contours are in units of mJy\,beam$^{-1}$, with a
    beam FWHM of 1.2\,arcmin. The contour levels are 0.1, 0.45, 0.8
    and 1.15\,mJy\,beam$^{-1}$.}
  \label{fig:amiconts}
\end{figure}

The measured fluxes from LDN\,1622 were compared to the expected AME
fluxes predicted by the spinning dust model and Interactive Data
Language (IDL) code \textsc{SpDust}
\citep{AliHaimoud2009,Silsbee2011}. The \textsc{SpDust} model expects
nine environment parameters in order to estimate the expected AME flux
from a region. The following is a brief description of each
environment parameter and its value: the hydrogen number density ($n_H
= 10^4$\,H\,cm$^{-3}$), the gas temperature ($T = 22$\,K), relative
intensity of interstellar radiation field ($\chi = 10^{-4}$), hydrogen
ionisation fraction ($x_H = 1$\,ppm), carbon ionisation fraction ($x_C
= 1$\,ppm), fractional abundance of molecular hydrogen ($y = 0$),
H$_2$ formation rate ($\gamma = 0$), rms of the dipole moment for dust
grains ($\beta = 9.3$) and the grain size distribution parameters
corresponding to a given line in \citet[Table 1, Line
7]{Weingartner2001}. A more detailed discussion of the \textsc{SpDust}
environment parameters can be found in \citet{AliHaimoud2009}.

\begin{figure*}
  \centering
  \subfloat{\includegraphics[width=0.3\textwidth]{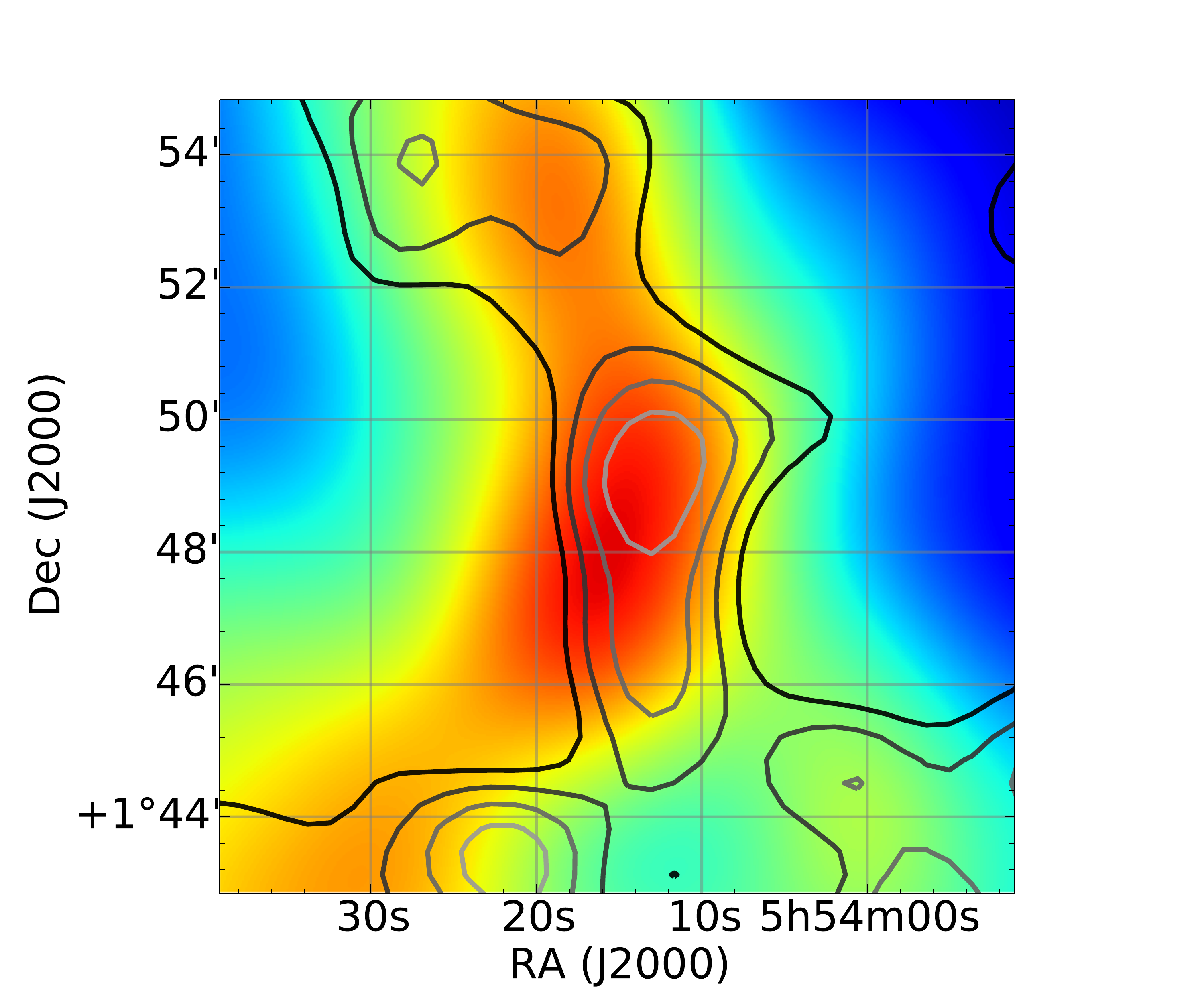}}
  \qquad
  \subfloat{\includegraphics[width=0.3\textwidth]{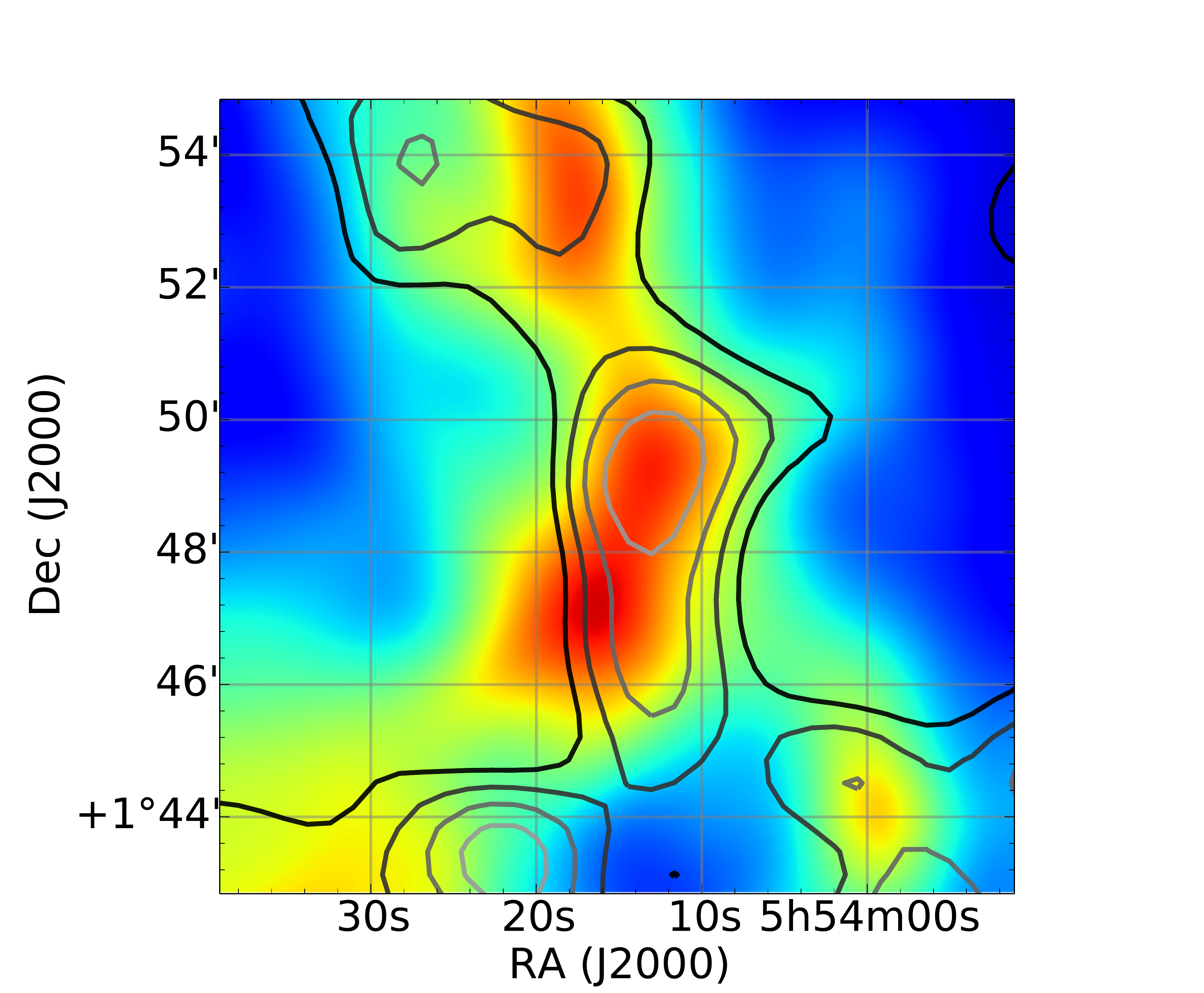}}
  \qquad
  \subfloat{\includegraphics[width=0.3\textwidth]{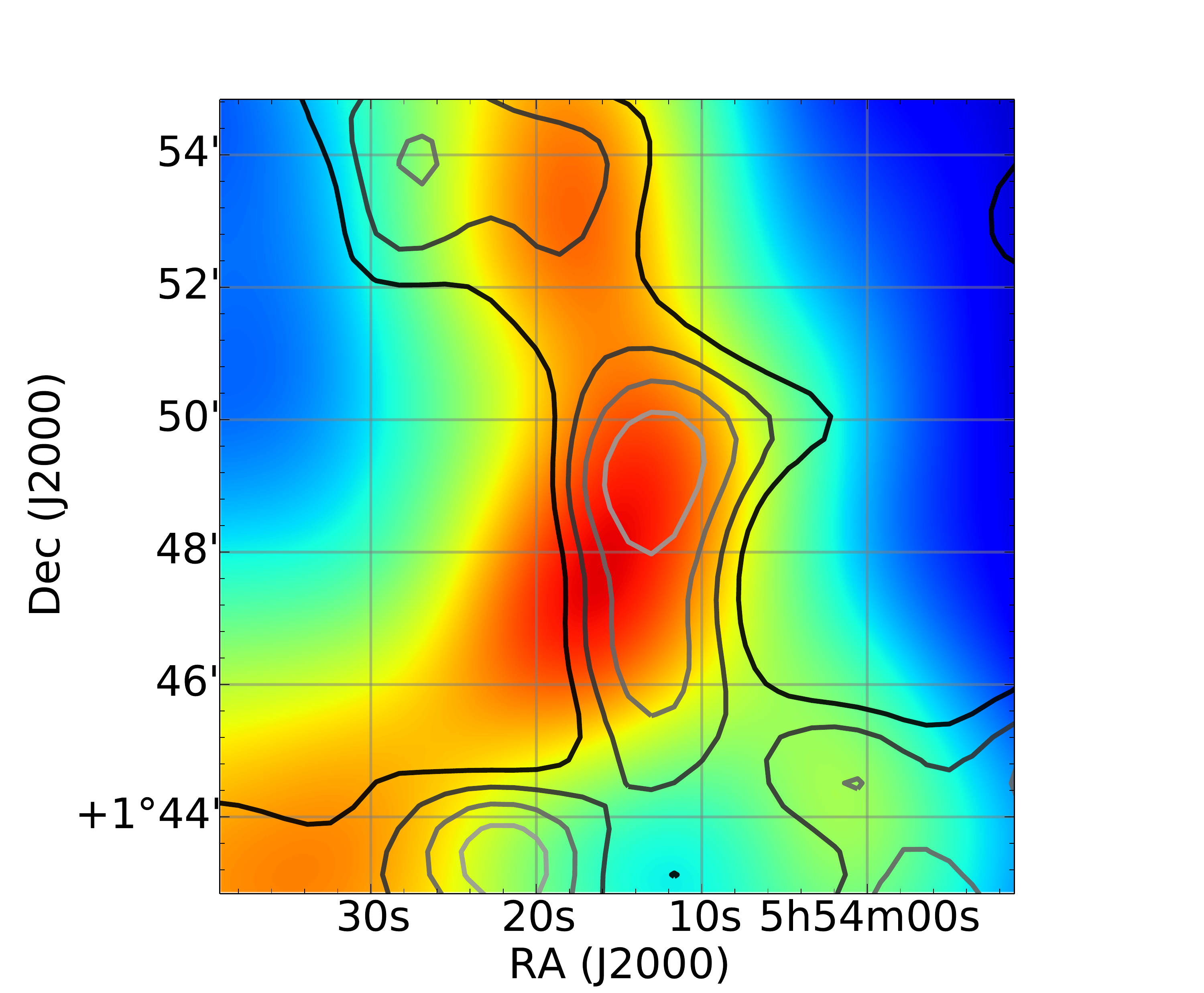}}

  \caption{ Overlay of smoothed Ku-band contours on the smoothed,
    point source removed MIR maps. \textit{Left} \textit{WISE}
    22\,$\mathrm{\mu}$m, \textit{centre} \textit{WISE}
    12\,$\mathrm{\mu}$m, \textit{right} \textit{Spitzer} IRAC
    8\,$\mathrm{\mu}$m. The contour levels are 0.1, 0.45, 0.8 and
    1.15\,mJy\,beam$^{-1}$ for a 1.2\,arcmin beam.}
  \label{fig:wiseconts}
\end{figure*}

In this \textit{Paper} the dark cloud environment parameters from
\citet{Draine1998b} are used. The expected hydrogen number density for
a dark cloud was checked against a calculated estimate of the true
hydrogen number density for LDN\,1622. The distance to LDN\,1622 is
typically considered to be similar to the Orion molecular clouds
($\approx$400\,pc) and it has an approximate angular size of
13\,arcmin. Assuming the cloud is spherical and using the measured
mean column density from \textit{Herschel} FIR data ($\approx 1.5
\times 10^{22}$\,H\,cm$^{-2}$), the estimated hydrogen number density
was found to be $\approx$3000\,H\,cm$^{-3}$, implying that the typical
dark cloud $n_H = 10^{4}$\,H\,cm$^{-3}$ is a reasonable order of
magnitude estimate for LDN\,1622. The gas temperature parameter was
matched to the mean dust temperature ($T_D$) within the Ku-band
aperture derived from the \textit{Herschel} FIR data, $T = T_{D} = 22\pm$2\,K.

A key indication of spinning dust is a rising spectrum between
$\approx$10\,$-$\,30\,GHz. The expected flux density within the
Ku-band aperture at 31\,GHz was estimated using the CBI contours shown
in \citet{Casassus2006} and scaling the flux density, assuming a
smooth source, to the expected flux within just the Ku-band
aperture. The spectral index between 13.7\,GHz and 31\,GHz was
measured to be 2.3$\pm$0.6. The uncertainty on the spectral index
was estimated using:
\begin{equation}
  \sigma_{\alpha} = \frac{1}{\mathrm{log}(\frac{\nu_2}{\nu_1})} \sqrt{ \left( \frac{\sigma_1}{S_1} \right)^2 + \left( \frac{\sigma_2}{S_2} \right)^2 } ,
\end{equation}
where $\nu$ is a given frequency, $S$ is the flux density at a given
frequency and $\sigma$ is the measured uncertainty. The spectral index
derived from \textsc{SpDust} between the same frequencies is 1.84,
which is 2.6\,$\sigma$ from the measured spectral index. The strong
confirmation that the spectral flux density spectrum is rising
between 13.7\,GHz and 31\,GHz and the agreement with the rising
spectral index predicted by \textsc{SpDust} is a good indication that
spinning dust is present within LDN\,1622 in the region of the Ku-band
aperture.

A map of LDN\,1622 using the AMI SA at 15.7\,GHz was provided for the
purposes of visual comparison with the Ku-band
map. Fig. \ref{fig:amiconts} shows the AMI SA data overlaid with
Ku-band contours. As both observations are at similar frequencies, and
both observations are sensitive to similar large scale
structures there should be significant correlations between the AMI SA
and Ku-band maps. Both the map and contours in Fig. \ref{fig:amiconts}
share the same large-scale elongated \textit{North-South}
structure. The most significant difference between the map and
contours in Fig. \ref{fig:amiconts} is the small displacement of the
Ku-band GBT data towards the \textit{West}.

It was shown in the previous discussion that the flux at Ku-band can
be naturally accounted for by the spinning dust model and that there
is a clear rising spectrum between 13.7\,GHz and 31\,GHz indicative of
AME.  However, if spinning dust is the origin of the Ku-band emission
there should be significant correlations between Ku-band and MIR
emission. The spinning dust model predicts that PAH molecules, VSGs or
a combination of both are the source of AME. VSGs generate MIR
continuum emission and PAH molecules emit bright MIR emission lines at
12.7\,$\mu$m, 11.3\,$\mu$m, 8.6\,$\mu$m and 7.7\,$\mu$m
\citep{Tielens2008}. The \textit{WISE} 12\,$\mu$m and \textit{Spitzer}
8\,$\mu$m passbands encompass these four PAH emission lines. The
\textit{WISE} 22\,$\mu$m passband only contains MIR continuum emission
from the VSGs. Therefore, if PAH molecules contribute to the
generation of AME, there should be a larger morphological correlation
between the Ku-band and the 12\,$\mu$m and 8\,$\mu$m maps than with
the 22\,$\mu$m map. It should be noted that it has been suggested that
MIR emission in some HII regions and PDRs could be due to the
formation of second-generation BGs in these environments
\citep{Everett2010,Draine2011,Paladini2012}. There is a possibility that
second-generation BGs are contributing to the MIR emission inside the
PDR around LDN\,1622, especially as the measured dust temperature within
the PDR is quite warm, $T_D = 22$\,K. However, for this
\textit{Paper} BGs were assumed to not be contributing to the
observed MIR emission.

In order to correlate the diffuse emission at Ku-band with the diffuse
emission in the MIR, the MIR maps had to first be filtered of point
sources. For both the \textit{WISE} and \textit{Spitzer} maps a number
of point sources near to the elongated \textit{North-South} structure were
patched out. The patching process involved replacing the source with a
2nd-order polynomial and noise estimated from the background around
the source. Due to the shorter wavelength and higher resolution of the
\textit{Spitzer} map there were a much larger number of weaker visible point
sources, which were then removed using a median filter. 

The patching and median filtering of the point sources left a number
of artefacts in the diffuse structure of the MIR maps at scales
comparable to the FWHM resolutions of the MIR maps. However, it is reasonable
to assume that these artefacts had minimal effect on this analysis as all
the MIR maps were smoothed considerably to match the 1.2\,arcmin FWHM
resolution of the smoothed Ku-band map.

The 1.2\,arcmin FHWM, point source filtered \textit{WISE} and
\textit{Spitzer} maps are shown in Fig. \ref{fig:wiseconts} with
Ku-band contours overlaid. All three maps and the contours share the
similar elongated \textit{North-South} morphology. All the maps in
Fig. \ref{fig:wiseconts} reveal that the peak in MIR emission also
coincides with the location of the peak Ku-band emission. From
visual comparison of the MIR maps with the Ku-band contours there is
no clear indication that either the \textit{Spitzer} 8\,$\mu$m or
\textit{WISE} 12\,$\mu$m maps, which contain the PAH emission lines,
have stronger correlations with the Ku-band contours than the MIR
continuum emission seen in the \textit{WISE} 22\,$\mu$m map.

The Pearson's correlation coefficient, $r$, between the Ku-band map
and each MIR map was measured to quantify any morphological
correlations between the maps. The pixels in the maps were made
quasi-independent by matching the pixel size to the FWHM of the
maps. The Pearson correlations for each MIR map with the Ku-band map
were found to be: $r$(22\,$\mu$m) = 0.54$\pm$0.13, $r$(12\,$\mu$m) =
0.59$\pm$0.13 and $r$(8\,$\mu$m) = 0.49$\pm$0.13. The uncertainties in
$r$ were estimated by bootstrapping the data and using the Fisher's
$r$ to $z$ transform \citep{Fisher1915}. The Pearson correlation
measurements imply that there is a slightly higher correlation between
the Ku-band and the MIR maps containing PAH emission lines. However,
the uncertainties on the correlation coefficients reveal there is no
significantly higher correlation between the Ku-band map and any one
MIR map.

In summary this Section finds evidence to support the possibility that
spinning dust in the form of either VSGs or PAHs is generating AME
within the PDR of LDN\,1622. The Ku-band emission measured by the GBT was
found to have little free-free contamination. A rising spectrum,
indicative of AME, was found between 13.7\,GHz and 31\,GHz. Finally,
morphological correlations were found between the Ku-band emission and
MIR emission.

\section{Conclusion} \label{sec:conclusion}
This \textit{Paper} has presented arcminute resolution mapping
observations of the diffuse radio emission associated with LDN\,1622
at C-band and Ku-band using the 100\,m GBT. The goal of the
observations was to measure free-free emission in the LDN\,1622 region
and identify and confirm the origin of the AME measured by previous
observations.

A free-free corona associated with the PDR of LDN\,1622 was revealed in
the C-band maps. The emission seen in the SHASSA H$\alpha$ map was
converted into the expected free-free brightness temperatures at
C-band and the peak brightnesses were found to agree within
$\approx$\,10\,\% of the peak observed C-band emission. However, much
of the the H$\alpha$ emission around LDN\,1622 was shown to be affected by
significant dust extinction. A simple model of the effect of dust
extinction on the H$\alpha$ emission was shown to be not sufficient to
correct for H$\alpha$ absorption over the whole cloud.

The Ku-band observations show a weak \textit{North-South} elongated
source inside the PDR of LDN\,1622. The diffuse nature of the Ku-band
emission made measurements of the flux density difficult. However, a
weak source at the core of an elongated structure was identified. The
flux density of the source was found to be consistent with previous
observations of LDN\,1622 and the flux density predicted by the
spinning dust model of AME. Evidence was found for a rising spectrum
between the measured GBT 13.7\,GHz emission and CBI 31\,GHz emission,
a strong indicator of spinning dust. Finally, the overall
morphological structure of the Ku-band emission was found to weakly
correlate with the emission from VSGs and PAH molecules at MIR
wavelengths. It should be acknowledged that the emission at Ku-band is
very weak and that the C-band observations do not entirely exclude the
possibility that the emission at Ku-band is at least partially
free-free emission. However, the correlations of the Ku-band emission
with MIR and ancillary radio data, along with the measured rising
spectrum of the source, do give reasonable confidence that the
observed 13.7\,GHz emission from LDN\,1622 is spinning dust driven
AME.


\section*{ACKNOWLEDGEMENTS}
SH acknowledges support from an STFC-funded studentship. CD
acknowledges support from an STFC consolidated grant (ST/L000768/1), an STFC Advanced Fellowship, an EU
Marie-Curie IRG grant under the FP7 and an ERC Starting (Consolidator)
Grant (no.~307209). The authors would like to thank Roberta Paladini,
Simon Casassus and Tim Pearson for helpful discussion and
comments. The authors would also like to thank Yvette Perrott and the
MRAO AMI group for providing the AMI SA observation of LDN\,1622.

\appendix 

\section{Maximum Likelihood Map-Making} \label{sec:AppA}

\begin{figure*}
  \centering
  \subfloat{\includegraphics[width=0.22\textwidth]{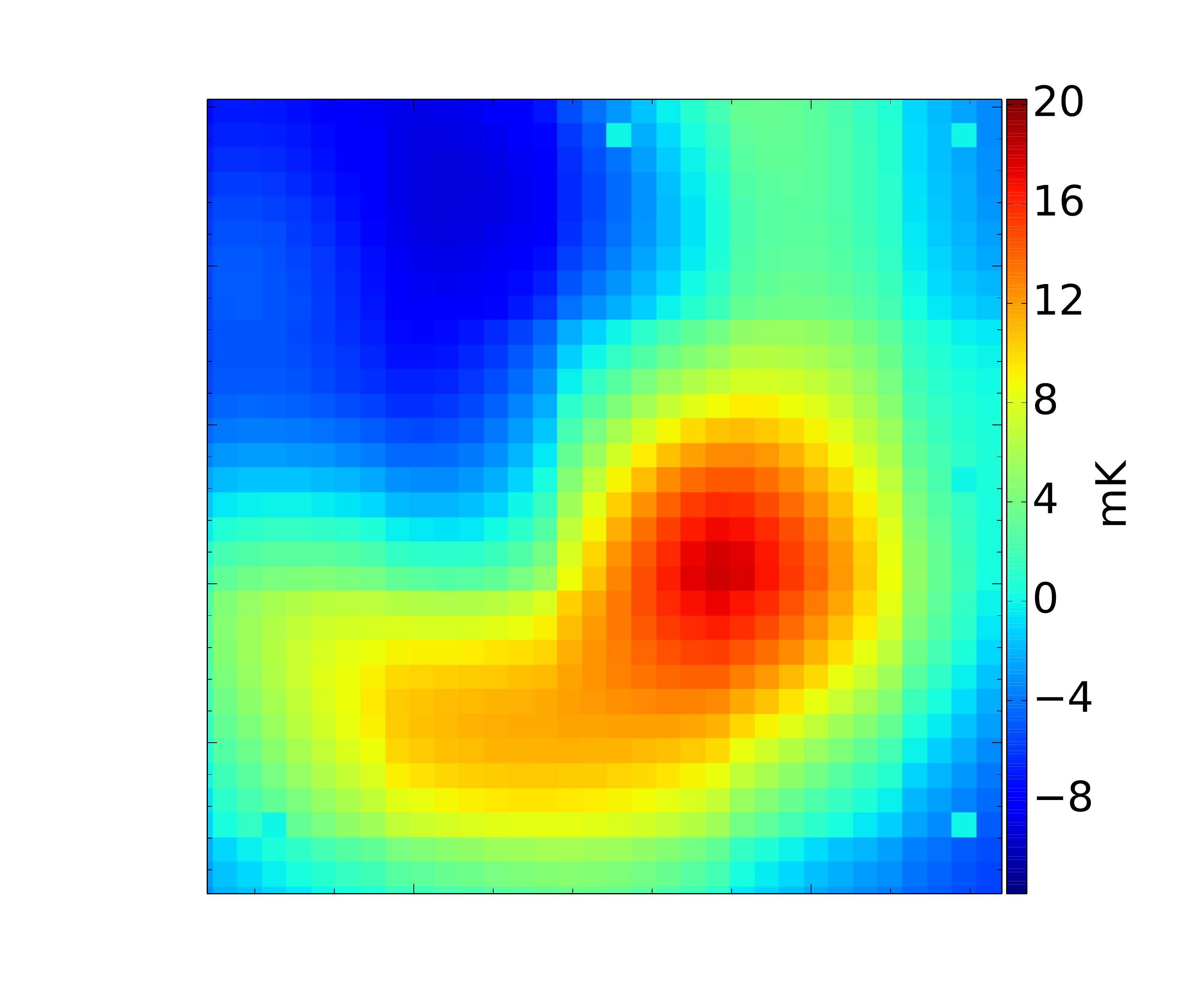}}
  \qquad
  \subfloat{\includegraphics[width=0.22\textwidth]{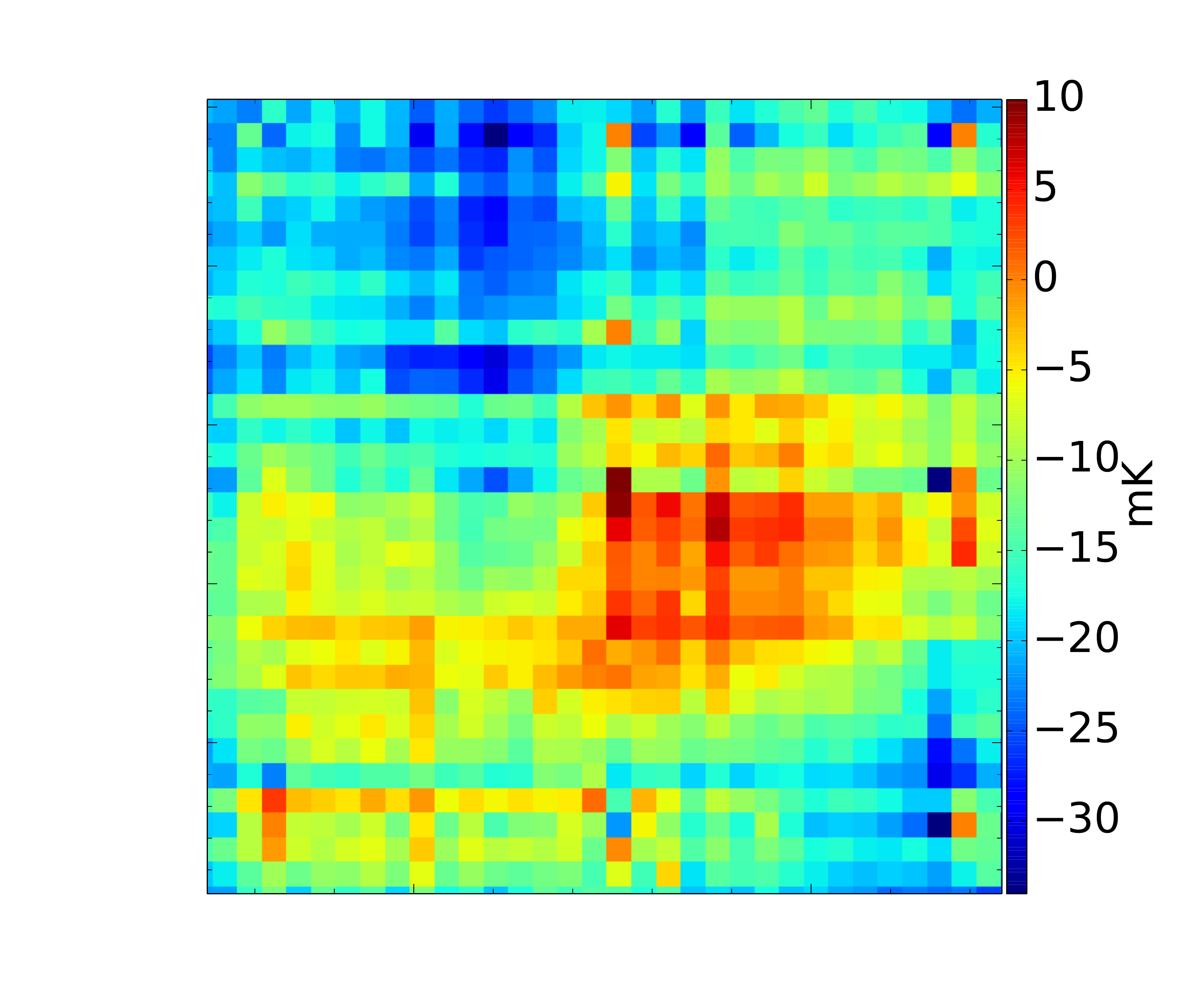}}
  \qquad
  \subfloat{\includegraphics[width=0.22\textwidth]{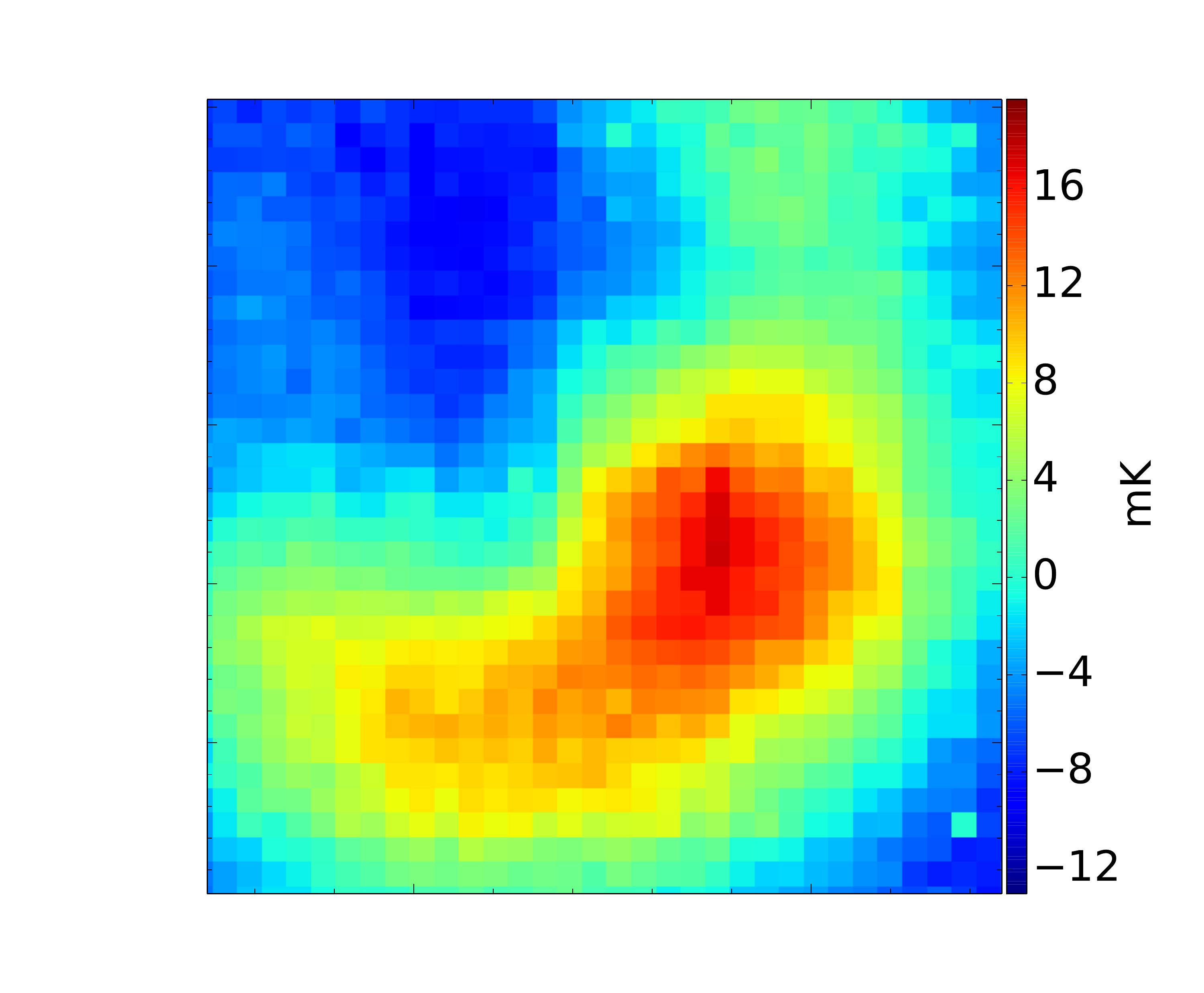}}
  \qquad
  \subfloat{\includegraphics[width=0.22\textwidth]{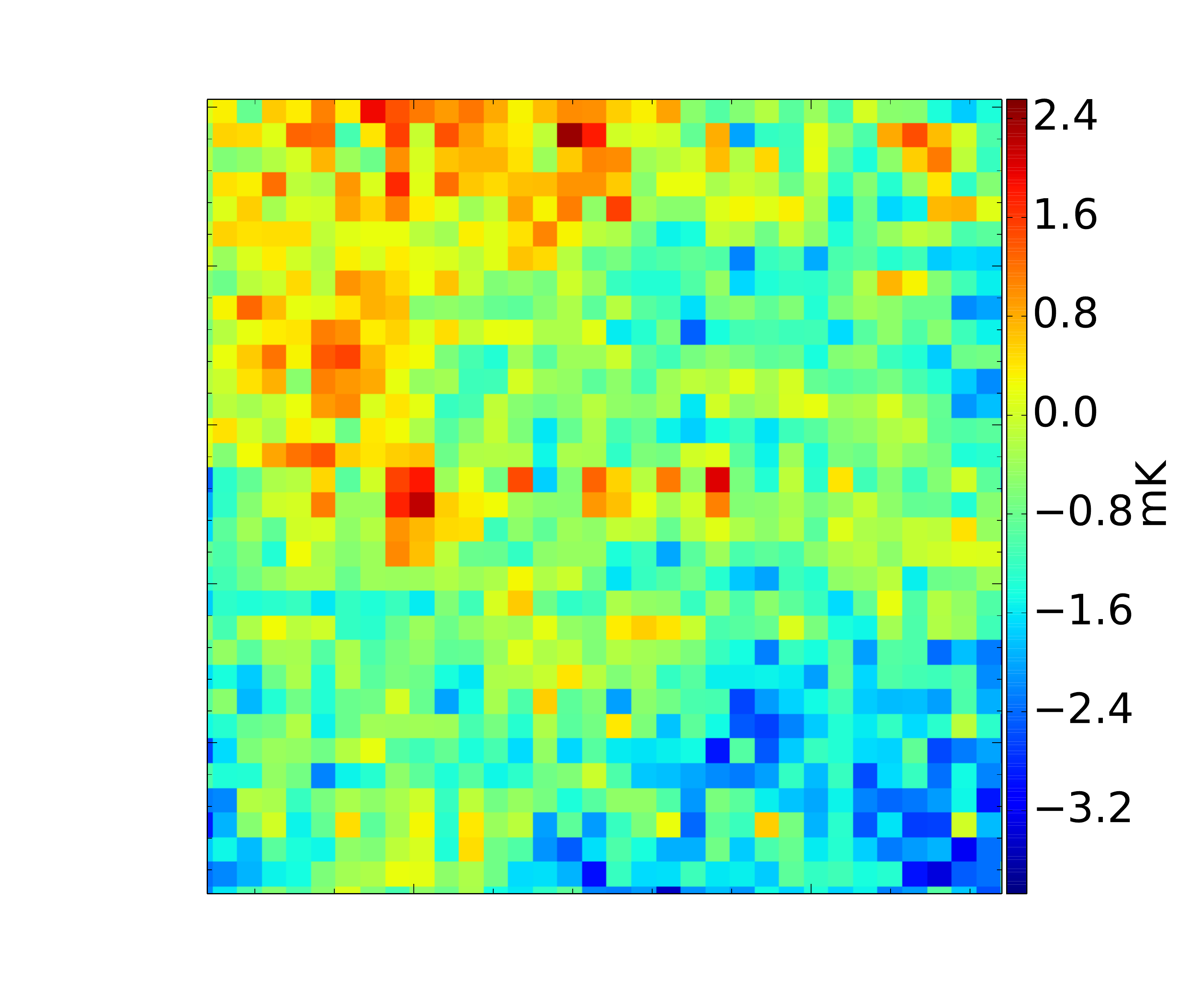}}

  \caption{ ML map-maker simulations. From \textit{Left} to
  \textit{Right} the maps show: noiseless input map; input map with
  \textit{1/f} noise contamination; recovered map using ML map-making;
  difference between the noiseless input map and recovered map.}
  \label{fig:simulations}
\end{figure*}

For this work an independent maximum-likelihood (ML) map-maker was
developed. The map-maker follows the methods outlined in previous ML
map-making papers
\citep{Borrill1999,Natoli2001,Dore2001}. The goal of ML map-making is
to iteratively solve for the full noise covariance matrix of the input
time-ordered-data (TOD) and remove stripes in the final map caused
by \textit{1/f} noise. The ML map-maker used in this \textit{Paper}
has been independently developed and implemented in
Python\footnotemark[2] with Python-callable FORTRAN libraries,
compiled using
\textsc{f2py}\footnotemark[3]. This Appendix briefly discusses simulations 
used to test the capabilities of the ML map-maker to remove \textit{1/f} 
noise and recover a known input signal.

\footnotetext[2]{Python Software Foundation. Python Language Reference, version 2.7. Available at http://www.python.org}
\footnotetext[3]{http://sysbio.ioc.ee/projects/f2py2e/}

Fig. \ref{fig:simulations} shows an input map that was sampled to
generate simulated noiseless TOD. Correlated noise was added to the
TOD using the \textit{1/f} noise model
\begin{equation}
P_{\nu} = \left( \frac{\sigma}{\nu_s}\right)^2 \left[ 1 + \left(\frac{\nu_{knee}}{\nu} \right)^{\alpha} \right] , 
\end{equation}
where $\sigma$ is the receiver sensitivity, $\nu_s$ is the sample
rate, $\nu_{knee}$ is the knee frequency, $\alpha$ is spectral index
and $\nu$ is the frequency of a given spectral bin. The parameters for
the simulated noise were chosen to match the noise of the C-band
GBT data.

Fig. \ref{fig:simulations} shows how a simulated diffuse source
contaminated with \textit{1/f} can be reliably recovered using the ML
map-maker. The residual map shows only white-noise and a residual
dipole. Typically, the ML map-maker cannot constrain the absolute
background (monopole) and the dipole moment of the map as there is not
sufficient information on those scales. However, the effect of the
dipole can typically be ignored as only the smaller scale structures
within a map are of interest.

For real data, ML map-making cannot remove all the effects of
\textit{1/f} noise. This is because the noise in real data will typically be 
non-stationary meaning the $\sigma$, $\nu_{knee}$ and $\alpha$ can all
vary with time. The effect of non-stationary noise can be reduced by
generating ML maps from subsets of the data where the noise is
quasi-stationary. However, splitting the data up is not always
possible because the integration time per pixel would be too low or
constraints due to the scanning strategy used.

\begin{figure}
  \centering
  \subfloat{\includegraphics[width=0.5\textwidth]{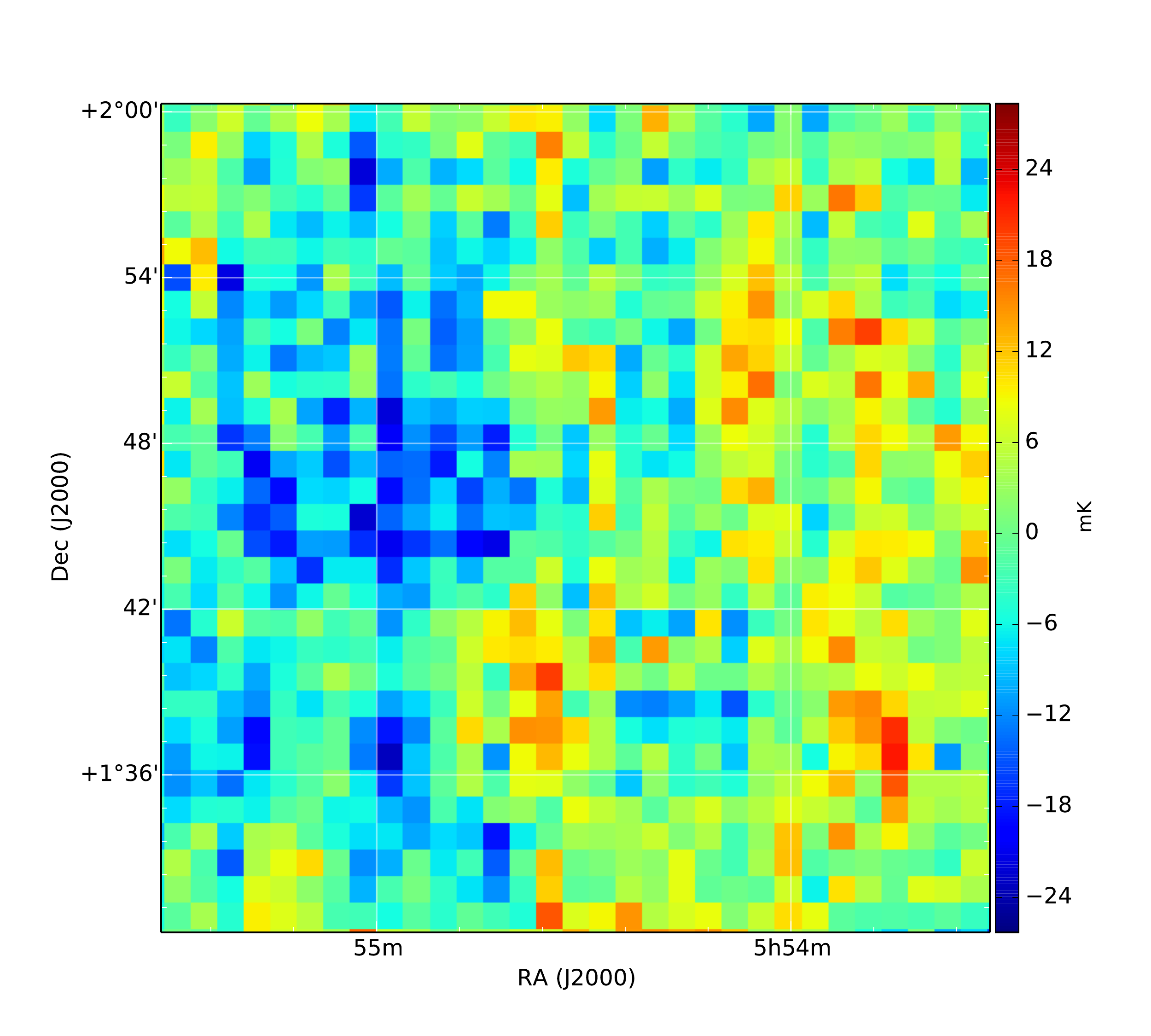}}

  \caption{Difference map of a jack-knife map generated from C-band
  data taken on the first day of observing and a jack-knife data
  generated from C-band data taken on the second and third days of
  observing.} \label{fig:JackKnife}
\end{figure}

In the case of the GBT data, the observations were split by
polarisation, as the receiver chains measured either LL or RR
polarised emission. Therefore, each polarisation would have slightly
different noise properties. A map for each polarisation was generated
and both were averaged to produce the final map. In
Fig. \ref{fig:JackKnife} a difference map generated from two
jack-knife maps of the C-band data is presented. Each jack-knife map
contained data from different days of observing. After ML map-making,
the two jack-knife maps were used to generate a difference map that
should contain statistically similar
\textit{1/f} noise as the C-band map.

Visually, Fig. \ref{fig:JackKnife} shows no clear evidence of the same
structures seen in the C-band data, similar to the difference map of the
simulated data in Fig. \ref{fig:simulations}. This implies that
there is no loss of signal when using ML map-making. However,
Fig. \ref{fig:JackKnife} does show evidence of correlated noise and a
gradient that have not been removed by the ML map-maker. This is due
to the noise, even when the data are split into days with similar
weather conditions, being only quasi-stationary. For this reason it is
expected that the sensitivity of the C-band and Ku-band maps presented
in this \textit{Paper} will still contain residual \textit{1/f} that
cannot be reliably removed without degrading the signal from the
astronomical source.

\bibliographystyle{mn2e}
\bibliography{References}

\end{document}